\documentclass[twocolumn]{aastex631}
\usepackage{amsmath}
\usepackage[T1]{fontenc}
\usepackage{ulem}

\newcommand\au{\mathrm{au}}
\newcommand\yr{\mathrm{yr}}

\newcommand\dy{~\mathrm{day}}
\newcommand\g{\mathrm{g}}

\newcommand\mol{\mathrm{mol}}

\begin{document}

\title{Formation of super-Earths and mini-Neptunes from rings of planetesimals}

\author{Sho Shibata}
\affiliation{Department of Earth, Environmental and Planetary Sciences, MS 126,  Rice University, Houston, TX 77005, USA}

\author{Andre Izidoro}
\affiliation{Department of Earth, Environmental and Planetary Sciences, MS 126,  Rice University, Houston, TX 77005, USA}




\begin{abstract}


The solar system planetary architecture has been proposed to be consistent with the terrestrial and giant planets forming from material rings at $\sim$1~au and $\sim$5~au, respectively. Here, we show that super-Earths and mini-Neptunes may share a similar formation pathway. In our simulations conducted with a disk $\alpha$-viscosity of $4\times10^{-3}$, super-Earths accrete from rings of rocky material in the inner disk, growing predominantly via planetesimal accretion. Mini-Neptunes primarily originate from rings located beyond the water snowline, forming via pebble accretion. Our simulations broadly match the period-ratio distribution, the intra-system size uniformity, and the planet multiplicity distribution of exoplanets. The radius valley constrains the typical total mass available for rocky planet formation to be less than 3-6 $M_\oplus$. Our results predict that planets at $\sim$1~au in systems with close-in super-Earths and mini-Neptunes are predominantly water-rich. Though relatively uncommon, at $\sim$1\% level, such systems might also host rocky Earth-sized planets in the habitable zone that underwent late giant impacts, akin to the Moon-forming event.

\end{abstract}



\section{Introduction} \label{sec: intro}

Super-Earths and mini-Neptunes, planets of sizes ranging from 1 to 4 Earth radii, orbit at least 30\% of sun-like stars in our galaxy~\citep{Howard+2010, Borucki+2010, Batalha+2013, Fressin+2013, Fabrycky+2014}. Their size distribution is bimodal, with peaks at approximately 1.4 and 2.4 Earth radii and a gap around 1.8 Earth radii, known as the ``exoplanet radius valley'' \citep{Fulton+2017, Fulton+2018}. This feature suggests a composition dichotomy: planets of about 1.4 Earth radii are likely dominated by silicate compositions, while those with sizes around 2.4 Earth radii may have significant volatile content in their cores or primordial H-He rich atmospheres~\citep{Kuchner03, Fortney+2007, Adams+2008, Rogers+2010, Owen+2017, Zeng+2019, Venturini+2020, Izidoro+2022}. Yet, the origins of these planets remain a subject of strong debate.

Several formation and post-formation mechanisms have been proposed to explain the exoplanet radius valley, including atmospheric photoevaporation \citep{Lopez+2013, Owen+2019}, core-powered mass loss \citep{Ginzburg+2018, Gupta+2022}, impact-induced mass loss \citep{Liu+2015, Inamdar+2016, Biersteker+2019, Kegerreis+2020, Matsumoto+2021, Chance+2022}, hydrogen sequestration into a magma ocean \citep{Kite2020}, and efficient hydrodynamic water mass loss \citep{Burn+2024}. Combining atmospheric mass loss mechanisms into planetary accretion models, some studies have broadly reproduced the observed radius valley \citep{Venturini+2020, Izidoro+2022, Burn+2024}. The effects of giant impacts \citep{Biersteker+2019} after gas disk dispersal are also consistent with the radius valley and additional constraints from observations as the period ratio distribution of planet pairs \citep[e.g.][]{Fabrycky+2014},  planet multiplicity distribution \citep[e.g.][]{lissaueretal11}, and the intra-system size uniformity \citep[e.g.][``peas-in-a-pod'']{Weiss+2018}. These accretion models, however, generally assume that planetesimals --  kilometer-sized building blocks of planets-- form across broad radial regions of the disk \citep[e.g.][]{Izidoro+2017, Izidoro+2022, Burn+2024}.


Planet formation starts from dust coagulation, drift, fragmentation, and turbulent mixing in a young protoplanetary disk \citep{Birnstiel+2024}. Planetesimals are thought to form under specific conditions where the local dust-to-gas ratio is sufficiently high for dust to concentrate into clumps and gravitationally collapse \citep{Youdin2005, Johansen+2007, Johansen+2009, Bai+2010}. Pressure bumps in the gaseous disk, often associated with condensation fronts~\citep{Bitsch+2015a, Hyodo+2019, Hyodo+2021, Charnoz+2021} or transitions in disk viscosity~\citep{Masset+06, Desch+2015, Flock+2017}, provide ideal sites for planetesimal formation~\citep{Johansen+2007, Taki+16, Drazkowska+2017, Lichtenberg+2021, Carrera+22}.

Recent planet formation models propose that the building blocks of solar system planets formed at pressure bumps near the silicate sublimation line and the water snowline \citep{Drazkowska+2018, Lichtenberg+2021, Izidoro+2021b, Morbidelli+2021}. As the disk's temperature evolved and the pressure bumps' location shifted, planetesimals were formed in ring-like distributions. The formation of the solar system's terrestrial planets is consistent with a ring of planetesimals extending from about 0.7 to 1.5 au, containing approximately 2 $M_\oplus$ of material~\citep{Hansen+2009, Walsh+2016}. The gas giant planets are thought to have originated from a second ring of planetesimals beyond the snowline~\citep{Brasser+2020, Lichtenberg+2021, Izidoro+2021b, Morbidelli+2021, Lau+2022, Lau+2024}, where more solid material was available for accretion, and pebble accretion was likely more efficient.
Asteroid populations and dynamical evolution models support the view that the solar system formed from rings of planetesimals rather than from a wide and continuous disk~\citep{Deienno+2024, Brasser2024}. 


In this work, we model planet formation under the scenario where planetesimals form in  ``rings'' at specific locations within the disk, rather than in continuous and widespread distributions \citep{Hansen+2009,Drazkowska+2017,Izidoro+2021b,Batygin+2023}. While previous models \citep[e.g.][]{Izidoro+2017,Izidoro+2022, Burn+2024} also propose either planetesimal accretion or pebble accretion as the primary mechanism for planetary growth, this study simultaneously accounts for both processes during planetary formation. In Sec.~\ref{sec: Methods}, we describe our formation model, which combines planetesimal accretion and pebble. We present the results of our numerical simulations in Sec.~\ref{sec: Result}. 
In Section~\ref{sec: Discussion}, we discuss our results, and in section \ref{sec: Summary} we present our main conclusions. 


\section{Methods}\label{sec: Methods}

\subsection{Code description}

We use N-body simulation modeling planetesimal accretion (mutual collisions between planetesimals), pebble accretion, gas-driven migration, and long-term evolution of planets starting from two rings of planetesimals. Our simulations are performed using the FLINSTONE code \citep{Izidoro+2022}, an adapted version of the MERCURY N-body integration package  \citep{Chambers1999}, which includes modules to compute the evolution of the protoplanetary disk, pebble accretion, and gas-driven migration,  \citep{Tanaka+2002, Paardekooper+2011}, and eccentricity and inclination damping~ \citep{Cresswell+2008}. 

We model the gaseous protoplanetary disk using 1D radial profile fits derived from hydrodynamical simulations \citep{Bitsch+2015b}. The disk $\alpha$-viscosity and disk metallicity are set to $0.004$ and $0.01$, respectively. Our disk model also features an inner disk cavity (inner edge). The structure of the cavity is modeled by rescaling the surface density at $\sim$0.1~au with a hyperbolic tangent function \citep{cossouetal14, Izidoro+2017}. The disk lifetime $t_\mathrm{disk}$ is an input parameter set as 2 or 3 Myr \citep[e.g.][]{williamscieza11}.

We model pebble accretion following \citet{Lambrechts+2014}. We set the size of silicate pebbles (inside the snowline, where T$<$ 170~K) as 1 mm in radius. This is consistent with dust evolution models and the typical sizes of chondrules observed in the solar system \citep{Guttler+2010, Birnstiel+2012, Friedrich+2015, Birnstiel+2024}. In our simulations, the total integrated pebble flux at the water ice line is $\sim280 M_\oplus$ when $t_\mathrm{disk}=2$ Myr, and $\sim320 M_\oplus$ when $t_\mathrm{disk}=3$ Myr, respectively. These values are consistent with those considered in previous studies \citep{Lambrechts+2019, Izidoro+2022}. We do not explore the effects of the pebble flux in our model. However, lower pebble fluxes would probably slow the growth of icy planets beyond the snowline and reduce their capacity to reach orbital periods smaller than 100 days during the gas disk lifetime \citep{Izidoro+2021a}. On the other hand, higher pebble flux would increase the number and masses of icy planets formed in the close-in orbit. These effects have been extensively explored in previous studies \citep[see][]{Izidoro+2021a}. The details of the disk evolution and pebble accretion models are described in Appendix~\ref{app: model}.

We track the composition of the accreted material of all planetary objects as they grow via impacts or pebble accretion. We assume that planetary materials inside the snowline have a silicate composition, and beyond the snowline, they contain a mixture of silicates and ice, with a water mass fraction of 50\%.

\subsection{Two-ring disk of planetesimals}

Our simulations model planet formation starting from two rings of planetesimals. We assume that planetesimals form during the very early stages of the disk when it is relatively hot. According to numerical simulations for solar-mass stars, during the build-up phase -- when the disk accretes material from its parent molecular cloud -- the silicate and water condensation lines can reach up to $\sim3$ au and $\sim17$ au, respectively. As the disk evolves and cools, these condensation lines then move inward \citep{Zhang+2015, Baillie+2019, Ueda+2021}. 
As pebbles coagulate in the disk~\citep{Birnstiel+2012}, they start to drift inward due to gas drag and can be concentrated at sublimation lines (of silicates and water), triggering planetesimal formation~\citep{Drazkowska+2018, Morbidelli+2021, Izidoro+2021b}. Numerical models investigating planetesimal formation show that the silicate and water condensation lines can form planetesimal rings around $0.2-1.5$ au and $3-15$ au, respectively \citep{Drazkowska+2017, Drazkowska+2018, Ueda+2021, Lichtenberg+2021, Izidoro+2021b}. The exact location where planetesimal rings form is not well constrained because it depends on the disk's temperature evolution, the structure of potential pressure bumps, and the pebble flux.
For simplicity, we set the inner and outer rings at $0.5-1.5~\au$ and $8-15~\au$, respectively, following the model by \citet{Izidoro+2021b}.

Formation models suggest planetesimals are born with typical sizes of about $\sim$100 km \citep{Johansen+2007, Simon+2016}. Planetesimals grow to larger objects via planetesimal-planetesimal accretion \citep[e.g.][]{Kokubo+2000} and/or pebble accretion \citep[e.g.][]{Johansen+2017}. N-body simulations modeling dynamical evolution and accretion starting from 100-km size objects are computationally prohibitive due to the large number of objects to carry the total disk mass. To reduce their computational cost, we start our simulations with Moon-mass planetary seeds (0.01-0.02 $M_\oplus$) in the inner ring and Ceres-mass planetary seeds (0.0005-0.001 $M_\oplus$) in the outer ring. These masses are broadly consistent with the masses that the largest planetary objects would have in these regions at about 300 kyr after the planetesimal formation. 


\begin{figure}
    \centering
    \includegraphics[width=0.9\linewidth]{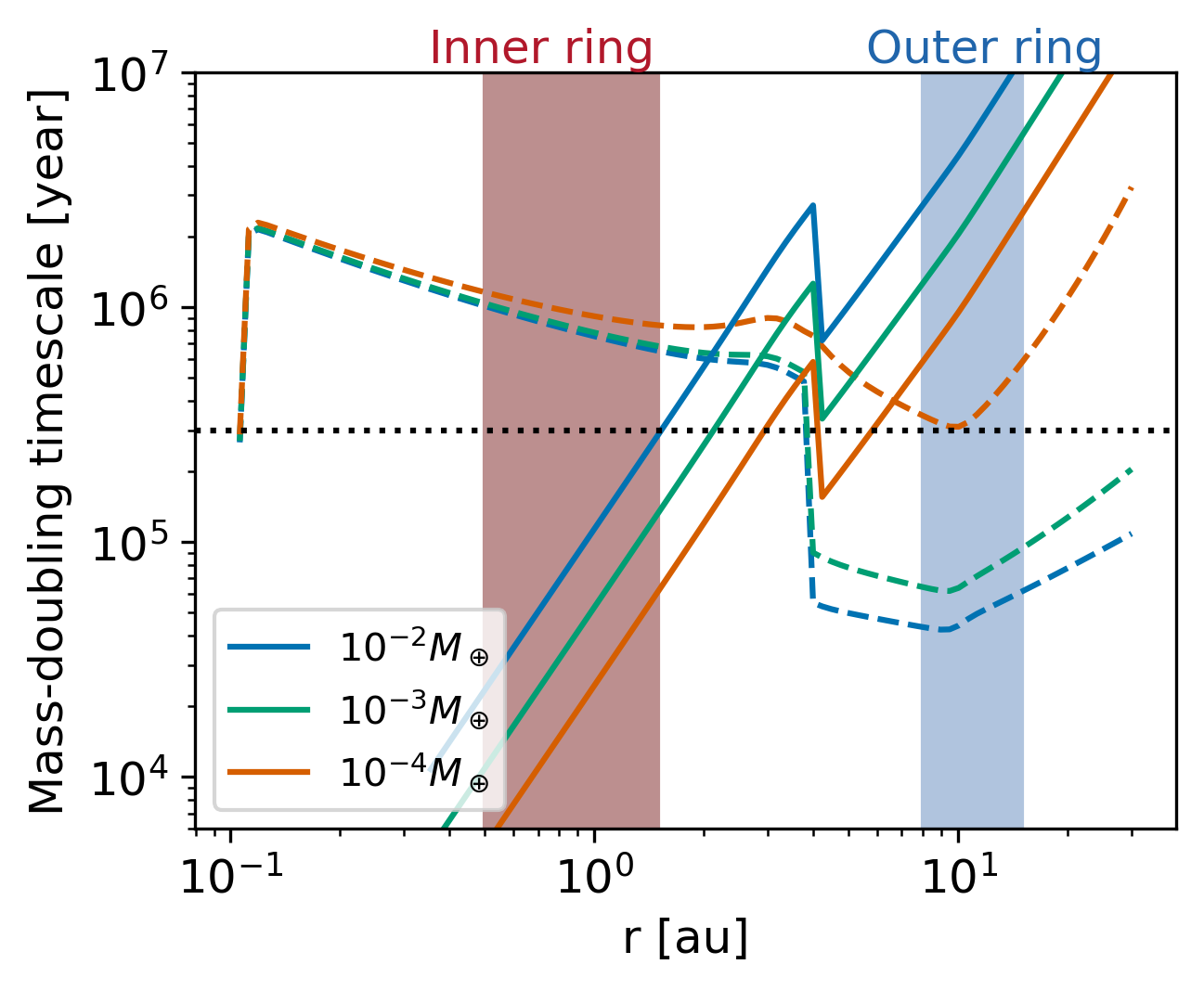}
    \caption{
    Mass-doubling timescale of seeds planets growing via planetesimal accretion (solid line) and pebble accretion (dashed line) across the disk. Different colors represent different planetary seed masses.
    The red and blue shaded areas are the inner and outer rings, respectively.
    The horizontal dotted line shows the time the simulation started. Timescales are computed at the beginning of the simulations.
    }
    \label{fig: r_timescale_growth}
\end{figure}

Figure~\ref{fig: r_timescale_growth} shows the mass-doubling timescale of planetary seeds at different disk locations. The location of the inner and outer rings are shown in red and blue, respectively. The solid lines show the planetesimal accretion timescale in the runaway growth regime. The surface density of planetesimals is assumed to be three times heavier than minimum mass solar nebulae \citep{Hayashi+1981}. We use a statistical model to compute the planetesimal growth timescale \citep{Inaba+2001, Fortier+2013} assuming planetesimals of 100 km in radius. The dashed line shows the pebble accretion timescale. The blue, green, and orange lines represent the growth timescales of planetary seeds of $10^{-2}$, $10^{-3}$, and $10^{-4} M_\oplus$, respectively.

In our model, the largest planetesimals growing in the inner ring via planetesimal accretion reach about Moon-mass ($10^{-2} M_\oplus$) within $\sim$300 kyr. Pebble accretion is inefficient in the inner ring because of the small size of silicate particles. In the outer ring, pebble accretion dominates planetesimal accretion due to the larger pebble sizes beyond the snowline. The largest planetesimals formed via gravitational collapse \citep{Klahr+2020} in the outer ring are expected to have sizes of up to $R\approx400$~km ($10^{-4} M_\oplus$). The largest planetesimals growing in the outer ring via pebble accretion are about Ceres-mass planetary seeds ($10^{-3} M_\oplus$) within $\sim$300 kyr.

The total number of seed planets in the inner ring depends on the initial ring mass, which is treated as a free parameter. Previous studies show that the total mass of a planetesimal ring is highly sensitive to the evolution of the gaseous disk, planetesimal formation efficiency, and dust grain sizes, which are not yet strongly constrained \cite[e.g.][]{Drazkowska+2017, Izidoro+2021b}. To explore a spectrum of possibilities, we perform simulations with different scenarios, considering inner rings ($0.5-1.5~\au$) with initial total mass in planetesimals equal to $M_\mathrm{disk} = 3 M_\oplus$, $6 M_\oplus$, $9 M_\oplus$, and $15 M_\oplus$. 

In the outer ring, we randomly distribute from 30 to 40 planetary seeds. Because planetesimal accretion is inefficient in the outer ring and the growth of the largest planetesimals beyond the snowline is driven by pebble accretion, we do not explore the total mass of planetesimals in the outer disk. The initial number of planetary seeds in the outer ring may appear relatively small. However, it is known that planetesimal and pebble accretion leads to oligarchic growth of objects, where the fastest growing objects tend to scatter smaller objects out of the disk pebble layer, preventing their growth~\citep{Levison+2015, Jiang+2022, Lau+2022}. In the end, only a relatively small number of oligarchs are formed.


\subsection{Simulation setup}


We performed eight sets of simulations exploring different combinations of $M_\mathrm{disk}$ and $t_\mathrm{disk}$ (four values for $M_\mathrm{disk}$ and two values for $t_\mathrm{disk}$). We label our different scenarios as {\it MxTx}. For instance, our scenario {\it M6T2} consists of an inner ring of $M_\mathrm{disk}=6 M_\oplus$ and disk dissipation timescale of $t_\mathrm{disk}=2$ Myrs. For each combination of parameters, we performed 50 simulations starting from slightly different initial distributions of planetary seeds. 
Our simulations start at $t=0.3$ Myrs after planetesimal formation and are integrated up to $t=50$ Myrs.

Previous studies have shown that most systems of super-Earths and mini-Neptunes experience dynamical instabilities within 20-50 Myr after disk dispersal \citep{Izidoro+2017}. 
While most of our simulations for $M_\mathrm{disk}=3 M_\oplus, 6 M_\oplus$, and $9 M_\oplus$ reached 50 Myrs, $\sim10\%$ of them did not reach 50 Myrs within a reasonable CPU time. As shown later (see Fig.~\ref{fig: Porb_Mp_for_one_param} in Sec.~\ref{sec: Result}), leftover seed planets often survive in the outer orbit at the end of the disk dissipation. The total mass of this leftover population is generally much smaller than that of the grown protoplanets, making their gravitational perturbations virtually negligible compared to the interactions between the protoplanets. However, despite their relatively low total mass, the presence of tens of leftover seed planets in the outer disk significantly slows down the numerical integration process. To accelerate this subset of simulations after 10 Myr of the integration, we removed leftover planetary seeds with masses smaller than $1\%$ of the biggest protoplanet. Notably, we ensure that the total mass of planetary seeds removed through this process is less than 3\% of the total mass of the planetary system itself. We found that removing this population does not prevent our inner systems from becoming dynamically unstable; for instance, neither affects the final architecture of our inner systems.

Our simulations with $M_\mathrm{disk}=15 M_\oplus$ were only integrated up to $t=20$ Myrs. In many of these simulations, planets at the disk's inner edge were pushed inside the cavity by the other inward migrating planets, requiring very short integration timesteps. Many of these simulations became computationally prohibitive. At 20 Myr, several of these simulations had already experienced orbital instability and formed too massive planets, inconsistent with super-Earth observations (e.g., compared to the observed radius distribution). Stopping them at 20 Myr does not affect the main conclusions of this study because planets formed with $M_\mathrm{disk}=15 M_\oplus$ does not fit with the observational data.

\subsection{Converting mass to planetary radius}

Our planet formation model provides planet mass, composition, and collision history. We follow the method used in \citet{Izidoro+2022} to calculate the planetary radius, which is based on mass radius relationships of \cite{Zeng+2019}. 


In our simulations, planets are systematically less massive than 10 $M_\oplus$ at the end of the gas disk phase. 
Both hydrodynamical simulations and analytic calculations show that their atmospheric masses would be limited to a few percent of their total masses \citep{Lee+2015, Ginzburg+2016}. Following previous studies \citep{Owen+2017, Ginzburg+2018, Gupta+2019, Izidoro+2022}, we assume that planets had atmospheric mass fractions $f_\mathrm{atm}$ of $0.003$ at the end of the gas disk phase. We show the cases where we use other values for $f_\mathrm{atm}$ in Appendix~\ref{app: mass_loss}.


If a planet experiences a giant impact ($M_\mathrm{p}/M_\mathrm{t}>0.1$: where $M_\mathrm{p}$ and $M_\mathrm{t}$ are the projectile to target masses) after the disk dissipation, we assume complete atmospheric loss due to the effects of the shock propagation and deposited thermal energy \citep{Liu+2015, Inamdar+2016, Biersteker+2019, Kegerreis+2020}. We also account for photoevaporation, assuming the energy-limited escape. We assume that a planet loses its primordial atmosphere if the received energy from the stellar irradiation is higher than the binding energy of the atmosphere. We follow a simple energy-limited escape prescription by \citet{Misener+2021}, which compares the atmospheric binding energy to the energy the planet receives from 100~Myr to 1 Gyr.


Following this approach, we group the obtained planets into four categories: rocky/icy core w/wo primordial atmosphere. Planets with water contents less/more than 10\% are categorized as rocky/icy cores.


\subsection{Observational biases}

We add observational bias to our planetary systems in order to effectively compare our results with observations. We simulate transit observations for each obtained planetary system and estimate the observational bias following the methodology described in \citet{Izidoro+2017, Izidoro+2022}, which accounts for geometric selection effects. During simulated observations, we also introduce random noise to the planetary radius, with a maximum value of 7\% of the planet's size, to account for systematics in observations \citep{Izidoro+2022}. 
The details of our method are provided in Appendix \ref{app: observational_bias}.

\section{Result}\label{sec: Result}
\subsection{Planet formation from two rings of planetesimals}

\begin{figure*}
    \centering
    \includegraphics[width=\linewidth]{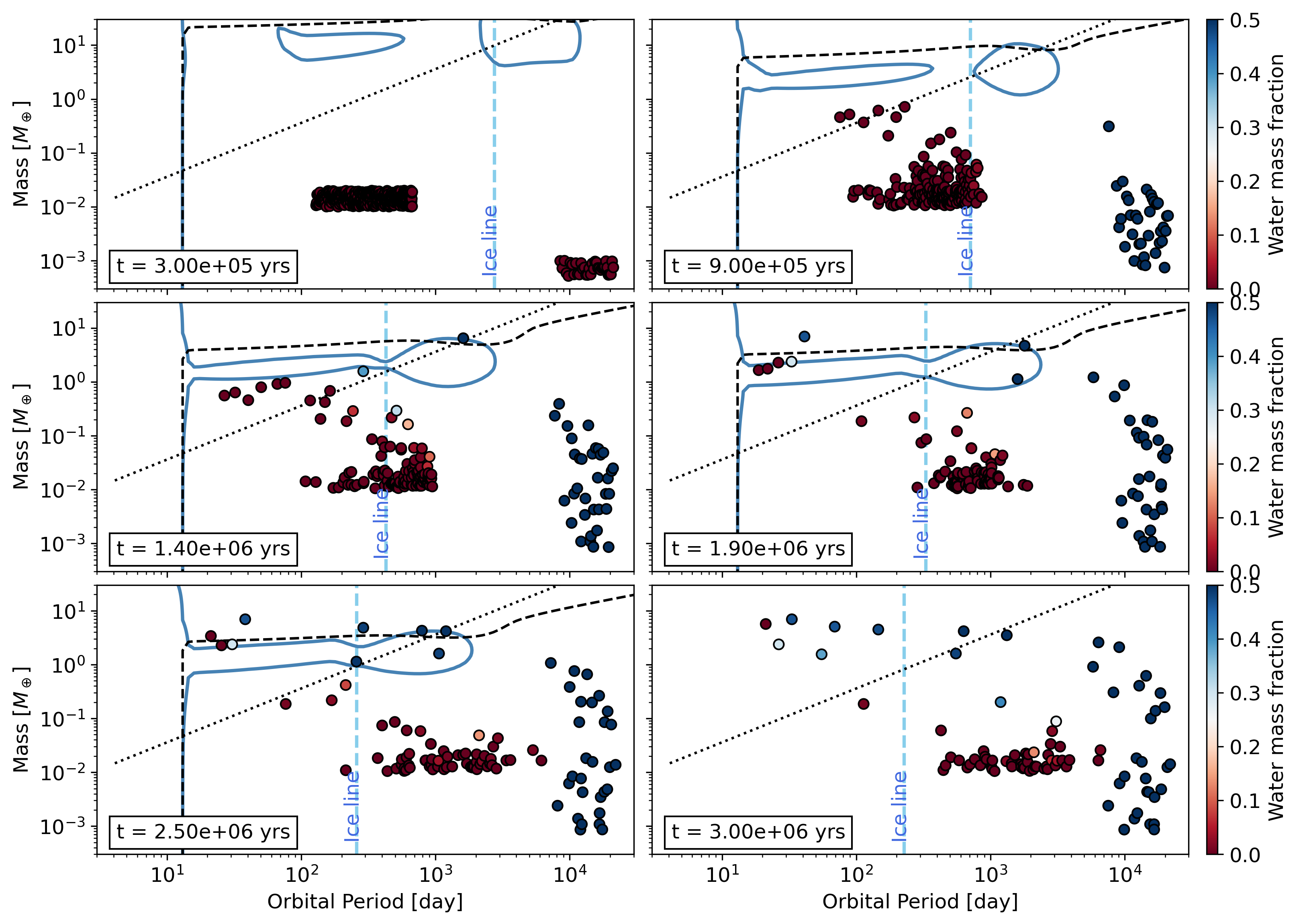}
    \caption{
    Snapshots show planetary seeds' growth and migration from two rings of material. This simulation comes from our {\it M6T2} scenario.
    The top-left panel shows the starting time of the simulation.   The time of each snapshot is shown on the bottom-left of each panel.
    The color of each dot corresponds to the water mass fraction of each planetary seed.  We plot planetesimal and pebble isolation masses with dotted and dashed lines, respectively.
    The planetesimal isolation mass is obtained by extrapolating the initial surface density of the inner planetesimal disk.
    The vertical blue dashed line shows the location of the water condensation line.
    The blue solid line is the boundary of the inward and outward migration where the migration torque is zero. 
    }
    \label{fig: Porb_Mp_for_one_param}
\end{figure*}

Figure \ref{fig: Porb_Mp_for_one_param} shows planetary seeds' growth and radial migration in one of our nominal simulations {\it M6T2}. Within the inner ring, planetary seeds primarily grow through planetesimal accretion. 
Although accounted for, pebble accretion is inefficient in the inner disk because of the small pebble sizes \citep{Johansen+2015}. Planetary seeds grow initially via mutual collisions by accreting objects in their local feeding zones. As planetary seeds grow, they migrate inward, reaching the disk's inner edge at about 0.1 au. During their inward migration phase, planetary seeds can suffer additional impacts, growing to masses beyond their original planetesimal isolation mass~\citep{Lissauer+1993}. 

Beyond the snowline, icy pebbles grow significantly larger, reaching sizes of up to a few centimeters \citep{Lambrechts+2014}. In the outer ring, pebble accretion dominates over planetesimal accretion because of the relatively larger pebble sizes. As icy planetary seeds grow via pebble accretion, they also migrate inwards. Inward migrating icy planets cross the inner ring region, dynamically scattering and accreting existing rocky planetary seeds. This process further depletes the 1~au region of rocky materials.

Our simulations show that as the disk evolves, the water snowline moves inward and sweeps through the inner planetesimal ring. This allows -- in some cases -- icy planets to form also from seeds originally in the inner ring. Convergent migration leads to the formation of resonant chains of multiple planets, including a mixture of planets originating from the inner and outer rings and carrying different water contents. 

Figure \ref{fig: Porb_Mp_for_one_param} shows a representative system's growth and orbital evolution during the gas disk phase. The subsequent evolution of these systems plays an important role in their final architecture. After gas disk dispersal, most systems  -- up to $\sim$90-95\% of our simulations --  become dynamically unstable, experiencing a phase of giant impacts and orbital reconfiguration characterized by the disruption of resonant chains \citep{Izidoro+2017}. We show the time evolution of mass and semi-major axis of planets in the simulation of Figure \ref{fig: Porb_Mp_for_one_param}  in Appendix~\ref{app: example_time_evolution}.

\subsection{Comparison with observations}
\begin{figure}
    \centering
    \includegraphics[width=\linewidth]{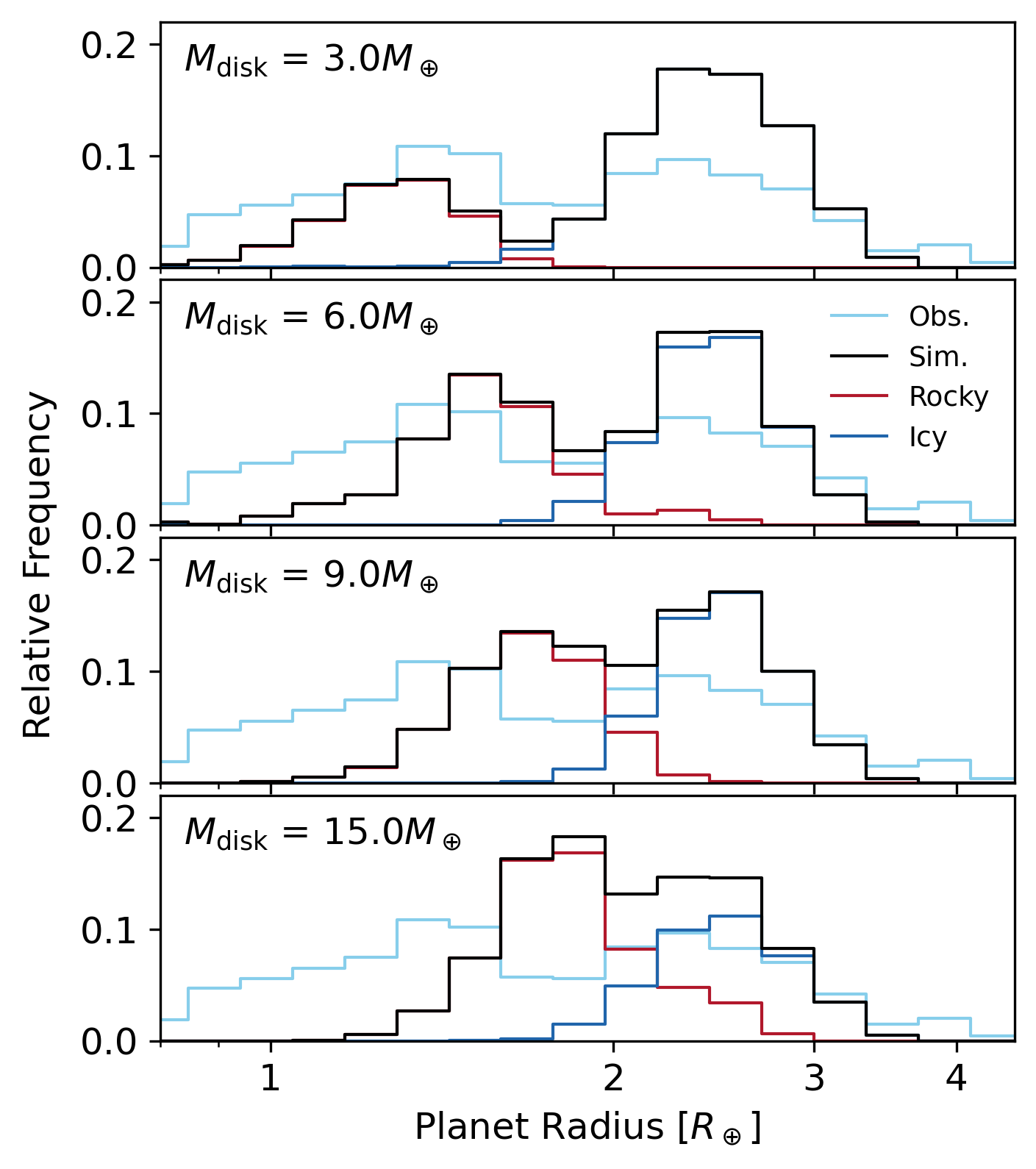}
    \caption{
    Radius distribution of planets at orbital periods shorter than 100 days.  The black line shows the results obtained in our simulations, including observational bias, and the light-blue line shows the observed planets from the California Kepler Survey. The red and blue lines show the rocky and icy planets, respectively. We define rocky planets as planets with a water mass fraction lower than $10 \%$. In these simulations, the disk lifetime is set to $t_\mathrm{disk}=3$ Myrs, and each panel shows the results of 50 simulations considering different inner ring masses as indicated at the top-left of each panel. }
    \label{fig: Radiut_Valley}
\end{figure}

We constrain our model using exoplanet observations from the California Kepler Survey (CKS). The CKS catalog contains about 907 planets with high-precision radius determination from 173 multi-planetary systems \citep{Fulton+2017, Fulton+2018}. To compare the results of our simulations with CKS planets, we selected planets with radii between 0.5 and $4.5 R_\oplus$ and orbital periods shorter than $100$ days, where observational completeness is relatively high~\citep{Fressin+2013}. 
We constrain our model using  radius~\citep{Fulton+2017, Fulton+2018, Petigura+2022}, period ratio~\citep{Fabrycky+2014}, size ratio~\citep{Weiss+2018}, and planet multiplicity distributions~\citep{Fressin+2013}. 


Figure~\ref{fig: Radiut_Valley} shows the radius distributions produced from simulations with different inner ring masses and observations. The gaseous disk lifetime in Figure~\ref{fig: Radiut_Valley} is $t_\mathrm{disk} = 3$ Myr. From top-to-bottom, the inner ring masses are 3, 6, 9, and 15 $M_\oplus$, respectively. In light blue, we show the radius distribution from the CKS sample, which peaks at 1.4 and 2.4~$R_\oplus$. The thin, dark blue, and red lines show the distribution of icy and rocky planets. In black, we show all the planets produced in each scenario.

The location of the radius valley in our model clearly depends on the initial inner ring mass. For $M_\mathrm{disk}=3M_\oplus$, the radius valley is slightly shifted relatively to the left, compared to the valley location of the CKS sample. For $M_\mathrm{disk}=6M_\oplus$, the location of the radius valley shifts towards the right, broadly matching observations. The valley moves further to the right for $M_\mathrm{disk}=9M_\oplus$, and the gap is ``filled up'' for $M_\mathrm{disk}=15M_\oplus$. 

Simulations considering shorter disk lifetimes of $t_\mathrm{disk} = 2$ Myr produced results broadly similar to those of Figure~\ref{fig: Radiut_Valley} where $t_\mathrm{disk} = 3$ Myr. The main difference is generally observed in the relative abundance of rocky and icy planets, namely the magnitude of the histogram peaks (see Appendix~\ref{app: all_hist}). In simulations with longer disk lifetimes, icy seeds have more time to grow by pebble accretion and migrate inwards. Consequently, the final relative number of icy planets inside 100 days in simulations considering longer disk lifetimes tends to increase. Previous planet formation models have suggested that planets at 1.4~$R_\oplus$ have rocky composition, whereas planets near the peak at 2.4~$R_\oplus$ are mainly icy planets \citep{Venturini+2020, Izidoro+2022}. Our radius distribution supports this hypothesis and constrains the typical planetesimal disk mass that super-Earths could have formed from to be less than $\sim$6$M_\oplus$.

\subsection{Mixing formation scenarios}\label{sec: discussion_mix_scenario}


The radius valley distributions of Figure \ref{fig: Radiut_Valley} indicate that inner rings having masses between 3 and 6 $M_\oplus$ provide a reasonable match to observations. It is unlikely, however, that all super-Earths and mini-Neptunes formed from rings of planetesimals with the same planetesimal mass and disk lifetime, but rather from a distribution of masses and lifetimes. To further constrain our model, we mix simulations of our different ring scenarios and find the best-fit mixing ratio to observations.


We look for the best-fit mixing ratio using the Kolmogorov-Smirnov test (KS). We calculate the KS statistic $\delta$ for each distribution: planet radius (size), period ratio of adjacent planet-pairs, size-ratio of adjacent planet-pairs, and planet multiplicity distributions. Our best-fit mixing ratio is defined as that with which the sum of all $\delta$ (for the four distributions) takes the minimum value. 
Appendix~\ref{app: test_mixing} provides an example illustrating how our method works. It is worth noting that our different scenarios lead to different levels of quality in fitting different observational constraints. For example, the size-ratio distribution of planets in the scenario {\it M3T3} provides a slightly better fit to observations than that from the {\it M6T2} scenario. 




\begin{figure}
    \centering
    \includegraphics[width=\linewidth]{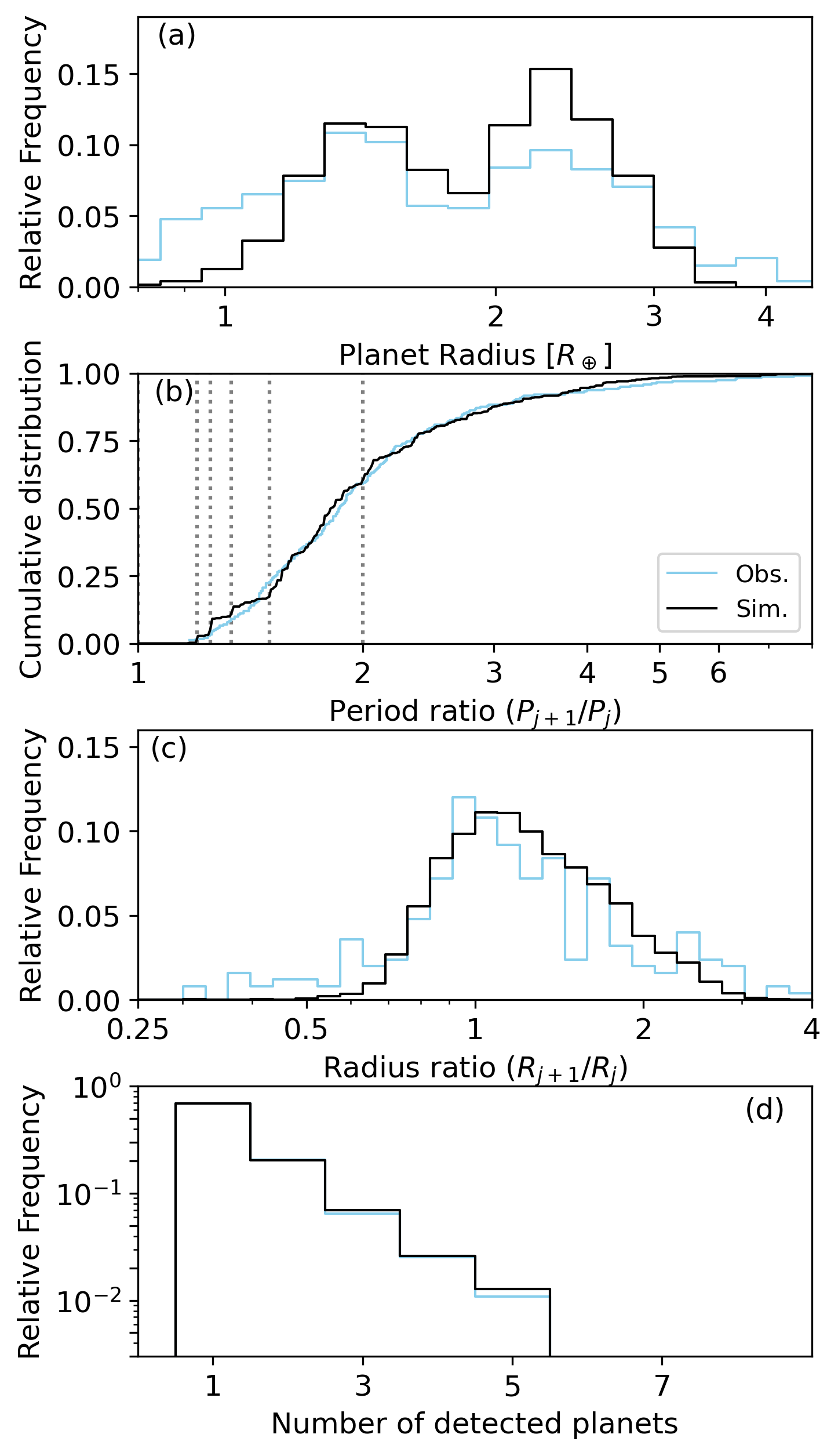}
    \caption{Comparison of the planetary architectures produced in simulations and exoplanet observations. The black line shows our simulations, mixing the planetary systems from our different models with the different weight fractions. Blue and black lines show observations (CKS) and the numerical results, respectively.
    The dashed lines in panels (b) and (c) show the histograms of all Kepler planets.
    {\bf Panel (a)}: radius distribution.
    {\bf Panel (b)}: distribution of the period ratio of adjacent planet pairs.
    The vertical dotted lines show the corresponding locations of 2:1, 3:2, 4:3, 5:4, and 6:5 mean motion resonances.
    {\bf Panel (c)}: distribution of the radius ratio of adjacent planet pairs.
    {\bf Panel (d)}: number of detected planets in each planetary system.
    }
    \label{fig: All_Mix}
\end{figure}



We found that combining 10\%, 20\%, and 70\% of systems from the M3T2, M3T3, and M6T2 scenarios provides the best fit to observations.
The distributions of the mixed systems are shown in Figure \ref{fig: All_Mix}. 
Our best fit to observations excludes cases with $M_\mathrm{disk}=9 M_\oplus$ (or larger masses). This result confirms our previous assertion that the typical rocky-mass reservoir in super-Earth and mini-Neptune systems is lower than $6 M_\oplus$.

\subsection{Radius valley for different ranges of orbital period}

\begin{figure}
    \centering
    \includegraphics[width=\linewidth]{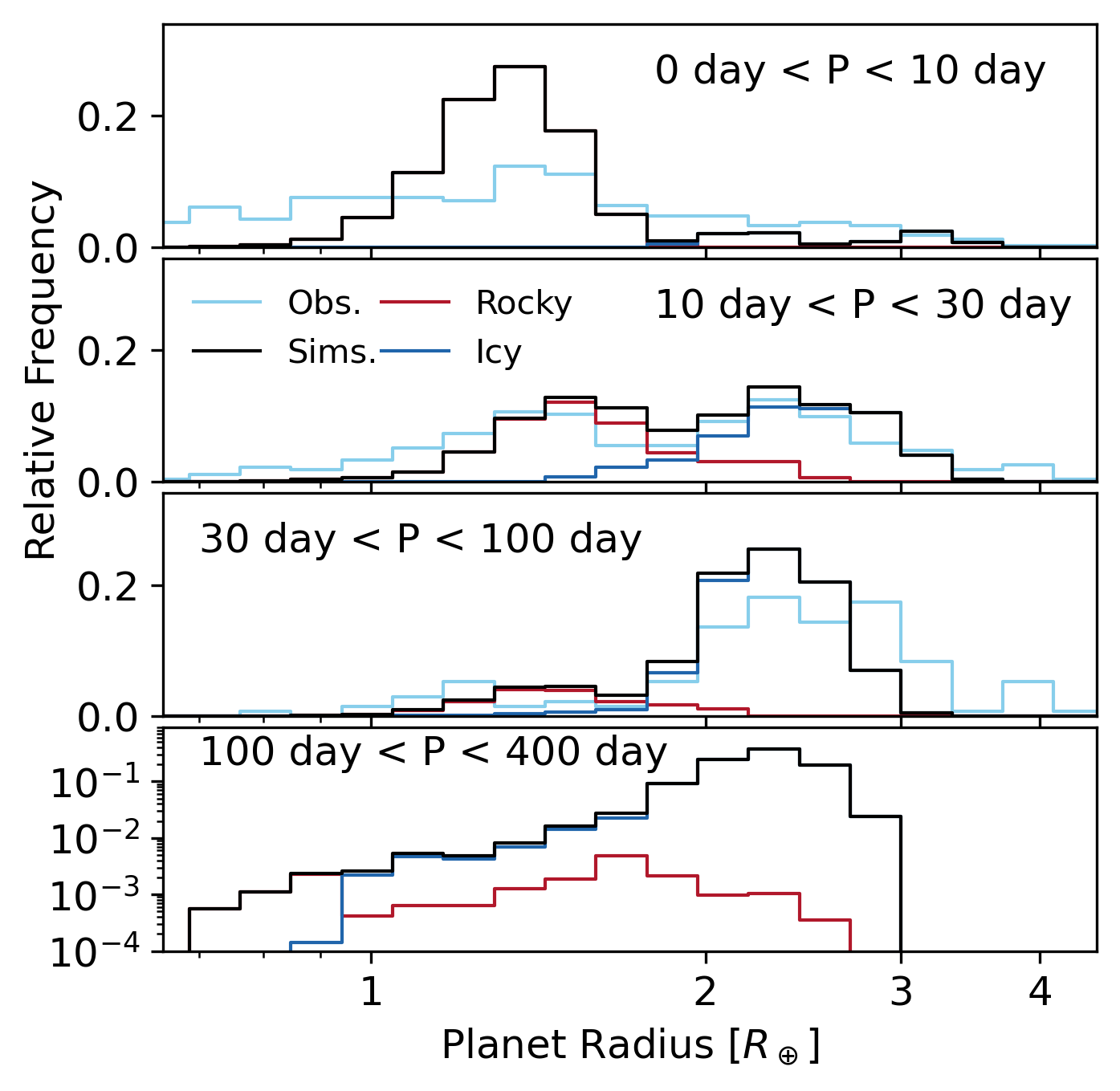}
    \caption{Planet radius distribution grouping planets into different ranges of orbital period. From top to bottom, panels show the radius distribution of planets with $0 \dy < P < 10 \dy$, $10 \dy < P < 30 \dy$,  $30 \dy < P < 100 \dy$, and $100 \dy < P < 400 \dy$. The dark-red and dark-blue solid lines show rocky and icy planets, respectively. The y-axis is the log scale only in the bottom panel.
    }
    \label{fig: Rhist_Porb}
\end{figure}

The planet radius distribution of the CKS sample also shows a dependence on orbital period, which is difficult to replicate in classical planet formation models invoking continuous distributions of planetesimals~\citep{Izidoro+2022, Burn+2024}. Figure~\ref{fig: Rhist_Porb} shows the radius distribution for three different ranges of the orbital period. Our results are broadly and qualitatively consistent with the distribution for all ranges of orbital periods. For instance, planets with orbital periods shorter than 10 days have mainly rocky compositions, generally having radii peaking at about 1.4 $R_{\oplus}$. Planets with orbital periods between 10 and 30 days show a bimodal distribution. Planets with orbital periods between 30 and 100 days are typically mini-Neptunes with sizes peaking at about 2.4 $R_{\oplus}$. In our model, planets at larger orbital periods generally come from the outer ring carrying significant amounts of water. Our model predicts that planets with orbital periods between 100 and 400 days are mainly mini-Neptunes, carrying high water contents. The bottom of Figure~\ref{fig: Rhist_Porb} shows that only about $\sim$1\% of our planets within this range of orbital periods have rocky compositions. 

\section{Discussion}\label{sec: Discussion}
\subsection{Statistical Comparison}\label{sec: discussion_Inconsistency}

We found that the best match to the observations is achieved by combining 10\%, 20\%, and 70\% of systems from the M3T2, M3T3, and M6T2 scenarios, respectively. To quantitatively evaluate how well our best-fit results align with observations, we performed a Kolmogorov-Smirnov (KS) test following the methodology of \citet{Izidoro+2017} (see Appendix \ref{app: test_mixing} for details). The resulting p-values for our best-fit solution are 0.05, 0.42, 0.35, and 0.88 for the radius, period ratio, radius ratio, and multiplicity distributions, respectively.


For the radius distribution, our KS test rejects the null hypothesis if we assume a significance level of $\sim95 \%$. The main reason for this mismatch between our results and observations is the lack of planets smaller than $1 R_\oplus$ in our models (see top panel of Figure \ref{fig: All_Mix}). When recalculating the p-value excluding planets smaller than $1 R_\oplus$, the p-value increases to 0.26. This statistical analysis hints that our model underestimates the population of small planets ($1 R_\oplus$). Although some small planets in the current data may be false positives, studies have suggested the existence of a substantial population of small sub-Earth planets \citep[e.g.,][]{qianwu20, Petigura+2022}. Consequently, this population offers additional constraints for formation models.

The final masses of planets growing in the inner ring depend on the planetesimal isolation mass, which scales with initial solid surface density and radial distance from the star. As expected, our simulations with the lowest initial inner ring mass (e.g.{\it M3T2}) form smaller rocky planets. The apparent lack of planets in our model with $<1 R_\oplus$ seems to suggest that some super-Earth systems (or sub-Earths) originate from inner planetesimal rings with masses lower than those we have considered in this study ($<$3$M_{\oplus}$), not so different from those considered in solar system formation models. This scenario will be investigated in a forthcoming paper.

For the other distributions (period ratio, size ratio, and multiplicity), our KS tests show no significant statistical differences between the simulated and observed populations. 
An interesting feature in the period ratio distribution is an apparent excess of planets with period ratios $P_\mathrm{j+1}/P_j < 1.5$, particularly near the 4:3 and 5:4 mean motion resonances (see panel (b) in Figure \ref{fig: All_Mix}). The p-value for this region is 0.01. Therefore, there is a statistical difference between the simulated and observed distributions, suggesting that our simulations overestimate the fraction of planet pairs in these resonances. However, this is not a critical issue in the model.

%
We stop our N-body simulations at 50 Myr. This is motivated by previous studies showing that most dynamical instabilities of resonant chains occur within $10$-$20$ Myrs after disk dispersal \citep{Izidoro+2017, Matsumoto+2021}. However, the long-term dynamical evolution of systems beyond our simulation time may still trigger additional orbital instabilities in resonant chains, further reducing the number of resonant planets. If we assume that some (but not all) resonant chains that survived in our simulations at 50  Myr experience orbital instabilities after 50 Myrs and deliberately remove the resonant chains from our analysis, the final period ratio distribution in Figure \ref{fig: All_Mix}-b gets closer alignment with observations. Previous studies have shown that the best match to the period ratio distribution of observations comes from a mixture of $\sim$1-10\% of systems that remain as resonant chains and $\sim$99-90\% of systems that experience dynamical instabilities (see \cite{Izidoro+2017, Izidoro+2021a} for a detailed discussion), depending on the specific formation model. A small fraction of stable resonant systems is necessary to account for resonant systems like Kepler-223~\citep{millsetal16}, TOI-178~\citep{leleuetal21} and Trappist-1~\citep{gillonetal17}.

It is also important to note that our model neglects several physical processes that could trigger additional instabilities, such as the effects of stellar tide \citep{bolmontetal16}, spin-orbit misalignment \citep{spaldingbatygin16}, and the presence of giant planets \citep[e.g.][]{bitschetal20}, etc. Our model also accounts for a single disk $\alpha$-viscosity and the disk viscosity may affect the final resonant state of systems \citep{bitschizidoro24}. All these processes and scenarios can play a role decreasing the number of planet pairs that survive in the 4:3 and 5:4 mean motion resonances. 

The apparent excess of resonant pairs at \( P_\mathrm{j+1}/P_j \lesssim 1.5 \) may also be an indirect consequence of the lack of small planets (\(\lesssim 1 R_{\oplus}\)) in our simulations.
After the disruption of resonant chains, the orbital separations of planet pairs are influenced by their combined mass, with more massive pairs tending to have wider separations \citep[e.g.,][]{Izidoro+2017}. Since our model does not produce relatively small planets (\(\lesssim 1 R_{\oplus}\)), planet pairs leaving resonances are scattered into wider separations and mainly contribute to the period ratio distribution at \( P_\mathrm{j+1}/P_j \gtrsim 1.5 \). In contrast, if a fraction of systems forms from relatively low-mass planetesimal rings (\(<3M_{\oplus}\)), more planet pairs may remain in compact configurations (\( P_\mathrm{j+1}/P_j \lesssim 1.5 \)) even after dynamical instabilities, potentially leading to a ``smoother'' period ratio distribution within \( P_\mathrm{j+1}/P_j \lesssim 1.5 \).




\subsection{A limited rocky reservoir}\label{sec: discussion_DiskMass}

Our results suggest that typical systems of super-Earths form from planetesimal rings carrying up to $3-6 M_\oplus$ of rocky material. 
Solar system formation models typically require $\sim$2-2.5 $M_\oplus$ in rocky material to explain the terrestrial planets~\citep{Hansen+2009, Walsh+2016}. Our results suggest that planetesimal rings forming super-Earths are about 2 to 3 times more massive than those forming the solar system's terrestrial planets. This implies that planetesimal formation in the inner ring of these systems was more efficient than in the solar system. It is interesting to note that meteorite data suggests that the inner and outer solar system were early separated into two disconnected reservoirs~\citep{Kruijer+17}, either by Jupiter's early formation or the effects of a pressure bump at the snowline~\citep{Brasser+2020, Izidoro+2021a}. In systems of typical super-Earths, it is likely that either this disconnection did not occur or was less efficient, allowing more massive planetesimal rings to form \citep{Izidoro+2021b}. 

In our simulations, pebbles inside the snowline have a silicate composition and are about 1 mm in size.  Disk turbulent viscosity and fragmentation velocity set pebbles sizes \citep{Birnstiel+2012}. In our nominal simulations, the disk viscosity is set to $4\times10^{-3}$, consistent with our silicate pebble sizes.  Lower viscosity disks would allow larger silicate pebbles in the inner disk, which could increase the efficiency of rocky pebble accretion in the inner ring. However, forming rocky planets exclusively through pebble accretion presents three significant challenges. First, for planetesimals to efficiently accrete pebbles, they must reach a sufficiently large size, which likely necessitates some degree of initial planetesimal accretion before pebble accretion becomes the dominant growth mechanism. \citep{Johansen+2017}. Second, the formation of rocky planets exclusively via pebble accretion may require special conditions, such as planetesimal formation to have occurred only inside the water snowline ~\citep{Izidoro+2021a}, which is not supported by current models \cite[e.g.][]{armitage16, Drazkowska+2018}. Third, if the accretion of silicate pebbles becomes efficient, rocky planets may probably grow so massive (several Earth masses) that the rocky planets may fill the radius valley \citep{Izidoro+2022} under the effects of giant impacts. Therefore, all these issues suggest that pebble accretion is unlikely to have been the main formation mechanism of rocky super-Earths \citep{Izidoro+2022, Batygin+2023}.

\subsection{Formation of Earth-like planets}\label{sec: discussion_EarthLike}

\begin{figure}
    \centering
    \includegraphics[width=\linewidth]{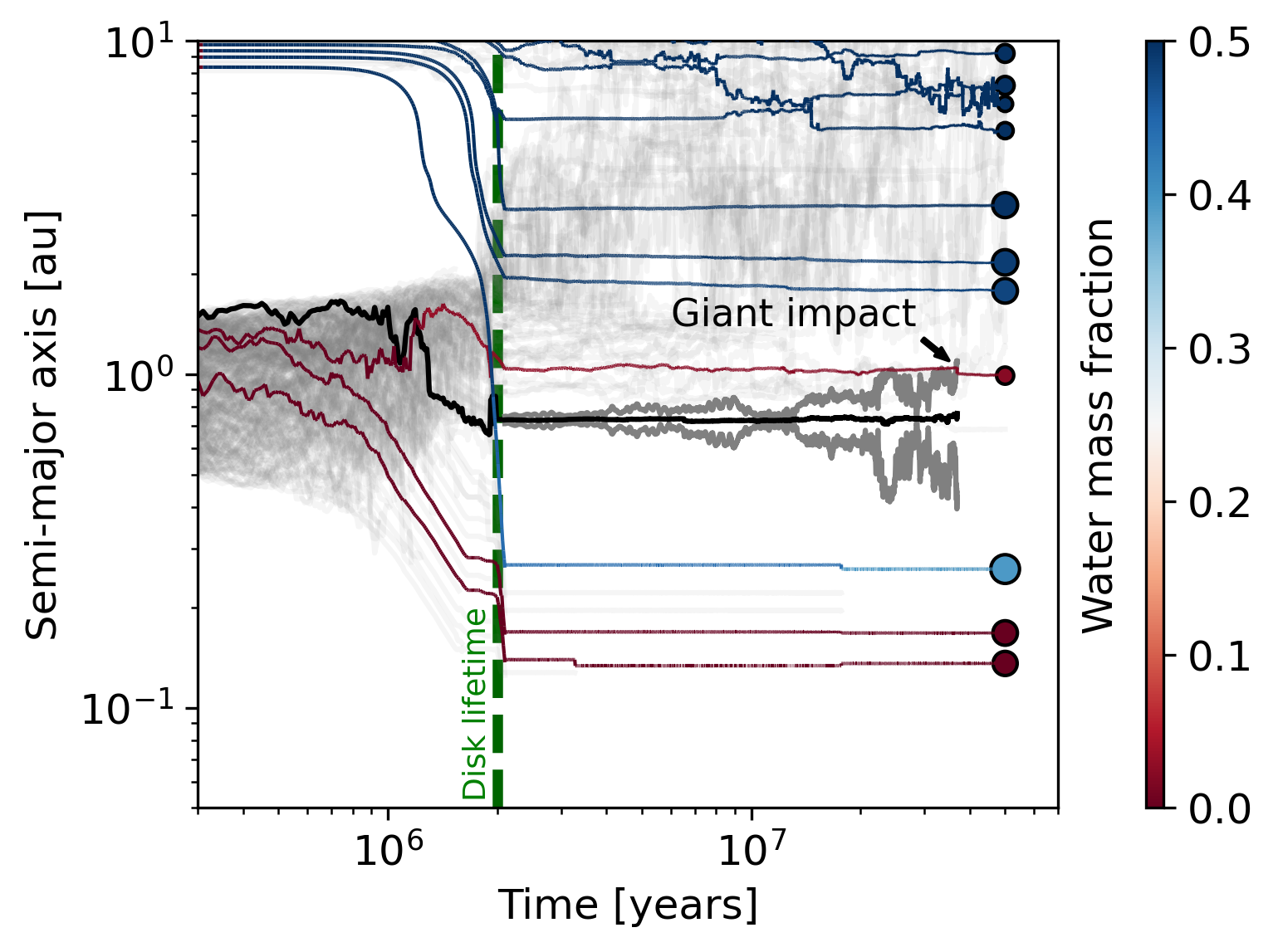}
    \caption{Dynamical evolution of a planetary system of close-in super-Earths and mini-Neptunes and an Earth-like planet at 1 au. 
    It shows the evolution of the semi-major axis of planets that survived at the end of the simulation in {\it M6T2}.
    The size of each dot at $t=50$ Myrs linearly scales with the planet's final radius, and the color corresponds to its water mass fraction.
    The rocky planet at around 1 au has a mass of $0.46 M_\oplus$ and experienced a giant impact with an object of about $0.04 M_\oplus$ at about 35 Myr. This is similar to the Moon-forming event in the solar system.
    The gray lines show the pericenter and apocenter of the impactor.
    The thin gray lines show the orbits of leftover planetary seeds. }
    \label{fig: time_axi_M6T2_45}
\end{figure}



Our formation model predicts that most planets with orbital periods between $100 \dy < P < 400 \dy$ are icy planets ($>10\%$ water content). This is because when icy planets migrate inward from the outer ring, they scatter and accrete rocky planets around $1 \au$. However, in a small fraction of our simulations, rocky Earth-like planets form around 1~au \citep{izidoroetal14}. Figure~\ref{fig: time_axi_M6T2_45} shows the formation history of our systems, which hosts a rocky planet of $0.46~M_\oplus$ at $0.99$ au carrying a water mass fraction of 2.2\%. For comparison, the total water content on Earth is estimated to be between 0.1\% and 0.5\% \citep[e.g.][]{peslier+2022}. Interestingly, this planet experienced a late giant impact at $t\sim3.5\times10^{7}$ yrs. The impactor-planet mass ratio is about 10\%, similar to the canonical model of the Moon-forming event \citep{Canup+2001}, constrained to have occurred between $\sim$30 and $\sim$150 Myr after the solar system's formation \citep{Yin+2002, Jacobsen+05, Allegre+2008}.


While the essential conditions for planetary habitability are not yet fully understood, taking our planet at face value, it may be reasonable to define Earth-like planets as those at around 1 AU with rocky-dominated compositions, protracted accretion, and relatively low water content. Our formation model suggests that such planets may also exist in systems of super-Earths and mini-Neptunes, although their overall occurrence rate seems to be reasonably low, about $\sim1\%$.

\section{Conclusion}\label{sec: Summary}
\begin{figure}
    \centering
    \includegraphics[width=\linewidth]{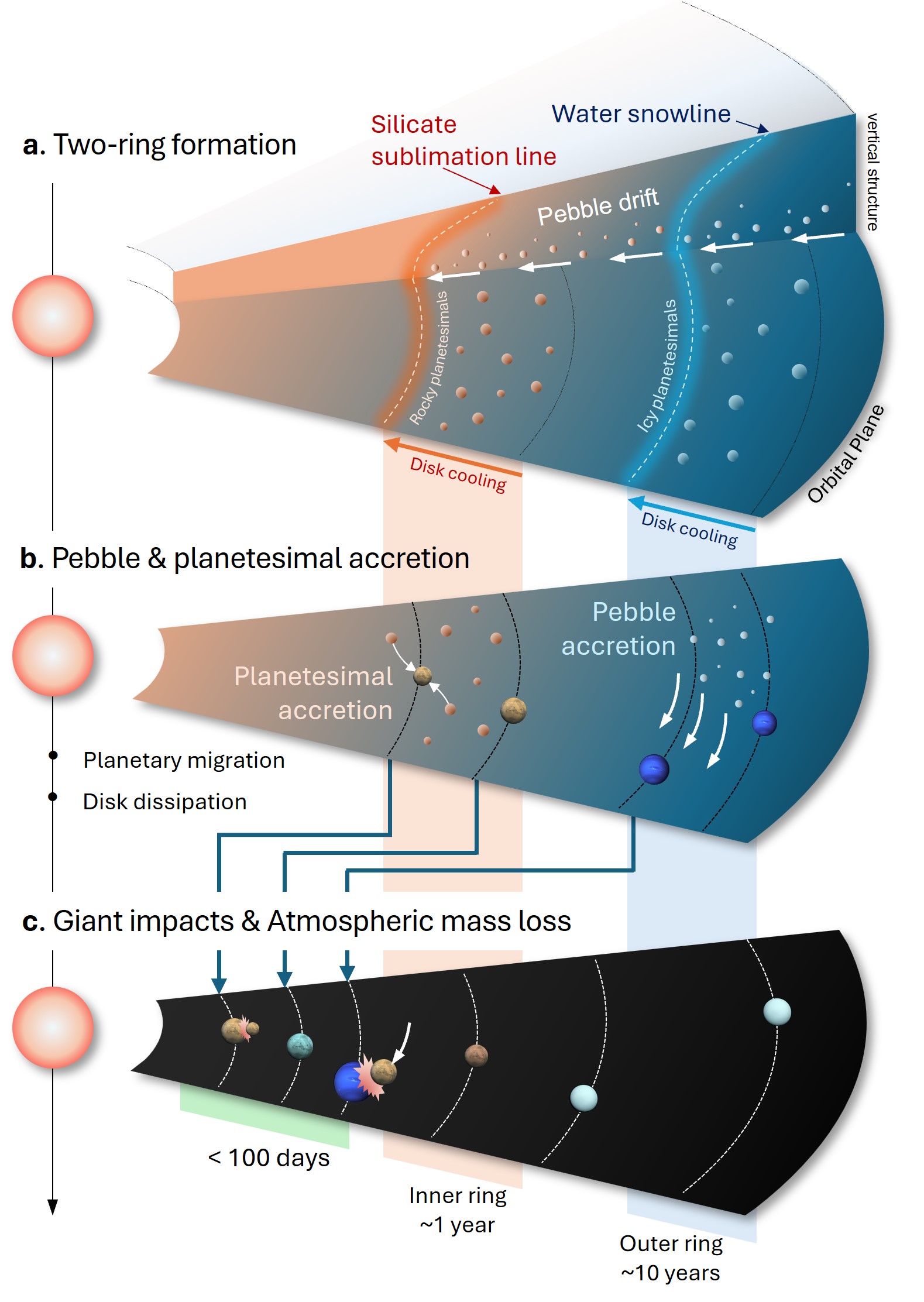}
    \caption{Schematic view of where and how super-Earths and mini-Neptunes form. Planetesimal formation occurs at different locations in the disk, associated with sublimation and condensation lines of silicates and water. Planetesimals and pebbles in the inner and outer rings have different compositions, as indicated by the different color coding (a). In the inner ring, planetesimal accretion dominates over pebble accretion, while in the outer ring, pebble accretion is relatively more efficient than planetesimal accretion (b). As planetesimals grow, they migrate inwards, forming resonant chains anchored at the disk's inner edge. After gas disk dispersal, resonant chains are broken, leading to giant impacts that sculpt planetary atmospheres and orbital reconfiguration (c). }
    \label{fig: two-ring-model}
\end{figure}

Our formation model demonstrates that the broad dynamical architecture of systems of super-Earths and mini-Neptunes is consistent with their formation from two rings of planetesimals. This differs from the classical picture where planets form from continuous distributions of planetesimals and is consistent with recent models of solar system formation \citep{Morbidelli+2021, Izidoro+2022}. Our model simultaneously accounts for the exoplanet radius valley, period ratio distribution, intra-system size uniformity, and planet multiplicity distribution of systems of super-Earths and mini-Neptunes. Our results suggest that super-Earths formed from an inner ring at around 1 au (inside the snowline). Mini-Neptunes are icy planets formed from a ring of planetesimals beyond the water snowline ($>$5-10~au). While super-Earths formed via the planetesimal accretion scenario, mini-Neptunes mainly grow via pebble accretion. Super-Earths and mini-Neptunes may have different formation sites and mechanisms (Fig.~\ref{fig: two-ring-model}).

We also provide insights on the occurrence rate and composition of planets between 100 and 400 days -- region overlapping with the habitable zone of sun-like stars~\citep{Kasting+1993} -- but poorly characterized by current observations~\citep{Batalha+2013, Fressin+2013, Rauer+2014}. Future observational surveys like PLATO \citep{Rauer+2014} will observe planets in this region and test our predictions. We also show that systems of super-Earths and mini-Neptunes may host rocky planets at around 1~au that experienced giant impacts after the gas disk dispersal, akin to the Moon-forming event~\citep{Canup+2001}.

\section*{Acknowledgments}
The authors are grateful for the constructive feedback from the anonymous referee.  S~S. and A.~I are grateful to the Center for Origins and Habitability of the Rice Space Institute. This work was supported in part by the Big-Data Private-Cloud Research Cyberinfrastructure MRI-award funded by NSF under grant CNS-1338099 and by Rice University's Center for Research Computing (CRC).
Part of the numerical computations were carried out on the Cray XC50 at the Center for Computational Astrophysics, National Astronomical Observatory of Japan.

%

\vspace{5mm}





\appendix

\section{Numerical model description}\label{app: model}

\subsection{Gas disk model}
We adopt 1D radial profiles for the gaseous disk. In the steady state condition, the surface density of gas $\Sigma_\mathrm{gas}$ is given as
\begin{equation}\label{eq: Sigma_gas}
    \Sigma_\mathrm{gas,0} = \frac{\dot{M}_\mathrm{gas}}{3 \pi \nu},
\end{equation}
where $\dot{M}_\mathrm{gas}$ and $\nu$ are disk accretion rate and the disk kinematic viscosity. Following \citet{Hartmann+1998} and \cite{Bitsch+2015a}, the disk accretion rate is given as
\begin{equation}\label{eq: disk_accretion_rate}
    \log_\mathrm{10} \left( \frac{\dot{M}_\mathrm{gas}}{M_\odot \mathrm{yr}^{-1}} \right) = -8 -1.4 \log_\mathrm{10} \left( \frac{t + 10^5 \yr}{10^6 \yr} \right),
\end{equation}
where $t$ is the age of the planetary system. We adopt the alpha-viscosity model \citep{Shakura+1973}, and the disk viscosity is given as 
\begin{equation}
    \nu = \alpha \Omega_\mathrm{K} {h_\mathrm{gas}}^2,
\end{equation}
where $\alpha$ is the dimensionless parameter, $\Omega_\mathrm{K}$ is the Keplerian orbital frequency, and $h_\mathrm{gas}$ is the scale height of the disk gas. For the ideal gas, the scale height of the disk gas is written as
\begin{equation}
    h_\mathrm{gas} = \frac{1}{\Omega_\mathrm{K}} \sqrt{\frac{\mathcal{R} T_\mathrm{disk}}{\mu}},
\end{equation}
where $T_\mathrm{disk}$ is the mid-plane temperature, $\mathcal{R}$ is the ideal gas constant, and $\mu$ is the mean molecular weight set to $2.3~\g~\mol^{-1}$. The mid-plane temperature is given by the fitting formulae provided in \citet{Bitsch+2015a}. The disk temperature depends on the disk metallicity $Z_\mathrm{disk}$ and $\alpha$. We set $Z_\mathrm{disk}=0.01$ and $\alpha=0.004$ in this study.

To account for the inner cavity, we introduce the cutoff function for the surface density of the disk gas, which is given as
\begin{equation}
    f_\mathrm{disk,in} = \tanh \left( \frac{r-r_\mathrm{in}}{0.05~\mathrm{au}} \right),
\end{equation}
where $r$ is the radial distance from the central star, and $r_\mathrm{in}$ is set to $0.1 \au$. Also, we introduce the exponential decay for the disk gas profile to account for the disk dissipation process. The decay factor $f_\mathrm{disk,diss}$ is given as
\begin{equation}\label{eq: disk_decay_factor}
    f_\mathrm{disk,diss} = 
        \begin{cases}
            \displaystyle{            
                1 
            } &{\rm for}~ t < t_\mathrm{diss,0}, \\
            \displaystyle{            
                \exp \left( -\frac{t-t_\mathrm{disk,0}}{t_\mathrm{disk}} \right),
            } &{\rm for}~ t_\mathrm{disk,0} < t, \\ 
        \end{cases}    
\end{equation}
where $t_\mathrm{disk,0}$ is set to $t_\mathrm{disk}-0.1\mathrm{Myrs}$. 

Finally, our disk model is given as
\begin{equation}
    \Sigma_\mathrm{gas} = f_\mathrm{disk,in} f_\mathrm{disk,diss} \Sigma_\mathrm{gas,0}
\end{equation}

\subsection{Pebble accretion model}
We adopt the pebble accretion model by \citet{Johansen+2017}. The pebble accretion rate is given by
\begin{equation}
    \dot{M}_\mathrm{p} = \pi {R_\mathrm{acc}}^2 \rho_\mathrm{p,mid} \Bar{S} \delta v,
\end{equation}
where $M_\mathrm{p}$ is the protoplanet's mass, $R_\mathrm{acc}$ is the accretion radius, $\rho_\mathrm{p,mid}$ is the midplane density of pebbles, $\Bar{S}$ is the stratification integral of pebbles, and $\delta v$ is the pebble's approaching speed given by $\Delta v +\Omega_\mathrm{K} R_\mathrm{acc}$ where $\Delta v$ is the sub-Keplerian speed. $R_\mathrm{acc}$ is obtained by solving the equation
\begin{equation}
    \tau_\mathrm{f} =\frac{\xi_\mathrm{B} \Delta v + \xi_\mathrm{H} \Omega_\mathrm{K} R_\mathrm{acc}}{\mathcal{G} M_\mathrm{p}/{R_\mathrm{acc}}^2},
\end{equation}
where $\tau_\mathrm{f}$ is the Stokes number, $\mathcal{G}$ is the gravitational constant, and $\xi_\mathrm{B}$ and $\xi_\mathrm{H}$ are fitting parameters. Numerical results are well reproduced with $\xi_\mathrm{B}=\xi_\mathrm{H}=0.25$ \citep{Ormel+2010, Lambrechts+2012}. The stratification integral $\Bar{S}$ is given as
\begin{equation}
    \Bar{S} = \frac{1}{\pi{R_\mathrm{acc}^2}} \int_{z_\mathrm{p}-R_\mathrm{acc}}^{z_\mathrm{p}+R_\mathrm{acc}} 2 \exp \left( -\frac{z^2}{2 {h_\mathrm{peb}}^2} \right) \sqrt{{R_\mathrm{acc}}^2-\left( z-z_\mathrm{p} \right)^2} \mathrm{d} z,
\end{equation}
where $z_\mathrm{p}$ is the height of the protoplanet measured from the disk's mid-plane, and $h_\mathrm{peb}$ is the scale height of the pebble layer.

The mid-plane density of pebbles is given as $\rho_\mathrm{p, mid}=\Sigma_\mathrm{peb}/\sqrt{2} h_\mathrm{peb}$ where $\Sigma_\mathrm{peb}$ is the surface density of pebbles, which is given by
\begin{equation}
    \Sigma_\mathrm{peb} = \frac{ f_\mathrm{peb} \dot{M}_\mathrm{peb}}{2 \pi r v_\mathrm{peb}},
\end{equation}
with
\begin{equation}
    v_\mathrm{peb} = \frac{2 \eta v_\mathrm{K}}{\tau_\mathrm{f}+{\tau_\mathrm{f}}^{-1}}+\frac{\nu}{r},
\end{equation}
where $f_\mathrm{peb}$ is the composition factor, $\dot{M}_\mathrm{peb}$ is the pebble flux, $v_\mathrm{peb}$ is the radial velocity of pebbles, and $v_\mathrm{K}$ is the Kepler velocity at the protoplanet's orbit. $f_\mathrm{peb}$ is set to 0.5 and 1.0 inside and outside the snowline, respectively. Inside the snowline, we calculate $v_\mathrm{peb}$ using the fixed pebble's size $s_\mathrm{peb}$, which is set to 1 mm in this study. Outside the snowline, the pebble's size is obtained by equating the pebble growth timescale with the drift timescale \citep{Lambrechts+2014}. In this case, the surface density of pebbles is transformed into
\begin{equation}
    \Sigma_\mathrm{peb} = \sqrt{\frac{2 \dot{M}_\mathrm{peb} \Sigma_\mathrm{gas}}{\sqrt{3} \pi \epsilon_\mathrm{p} r v_\mathrm{K}} },
\end{equation}
where $\epsilon_\mathrm{p}$ is the pebble's sticking efficiency set to $0.5$ in this study. 
The pebble flux $\dot{M}_\mathrm{peb}$ is given by the model developed by \citet{Lambrechts+2014}, given as
\begin{equation}
    \dot{M}_\mathrm{peb} = 2 \pi r_\mathrm{g} \frac{\mathrm{d} r_\mathrm{g}}{\mathrm{d} t} Z_\mathrm{disk} \Sigma_\mathrm{gas} (r_\mathrm{g})
\end{equation}
with
\begin{equation}
    r_\mathrm{g} = \left( \frac{3}{16} \right)^{1/3}  \left( \mathcal{G} M_\mathrm{s} \right)^{1/3} \left(\epsilon_\mathrm{D} Z_\mathrm{disk} \right)^{2/3} t^{2/3},
\end{equation}
\begin{equation}
    \frac{\mathrm{d} r_\mathrm{g}}{\mathrm{d} t} = \frac{2}{3} \left( \frac{3}{16} \right)^{1/3}  \left( \mathcal{G} M_\mathrm{s} \right)^{1/3} \left(\epsilon_\mathrm{D} Z_\mathrm{disk} \right)^{2/3} t^{-1/3},
\end{equation}
with $\epsilon_\mathrm{D}=0.05$.

\section{Atmospheric mass loss model}\label{app: mass_loss}

\begin{figure}[h]
\centering
\begin{minipage}[b]{0.49\columnwidth}
    \centering
    \includegraphics[width=\linewidth]{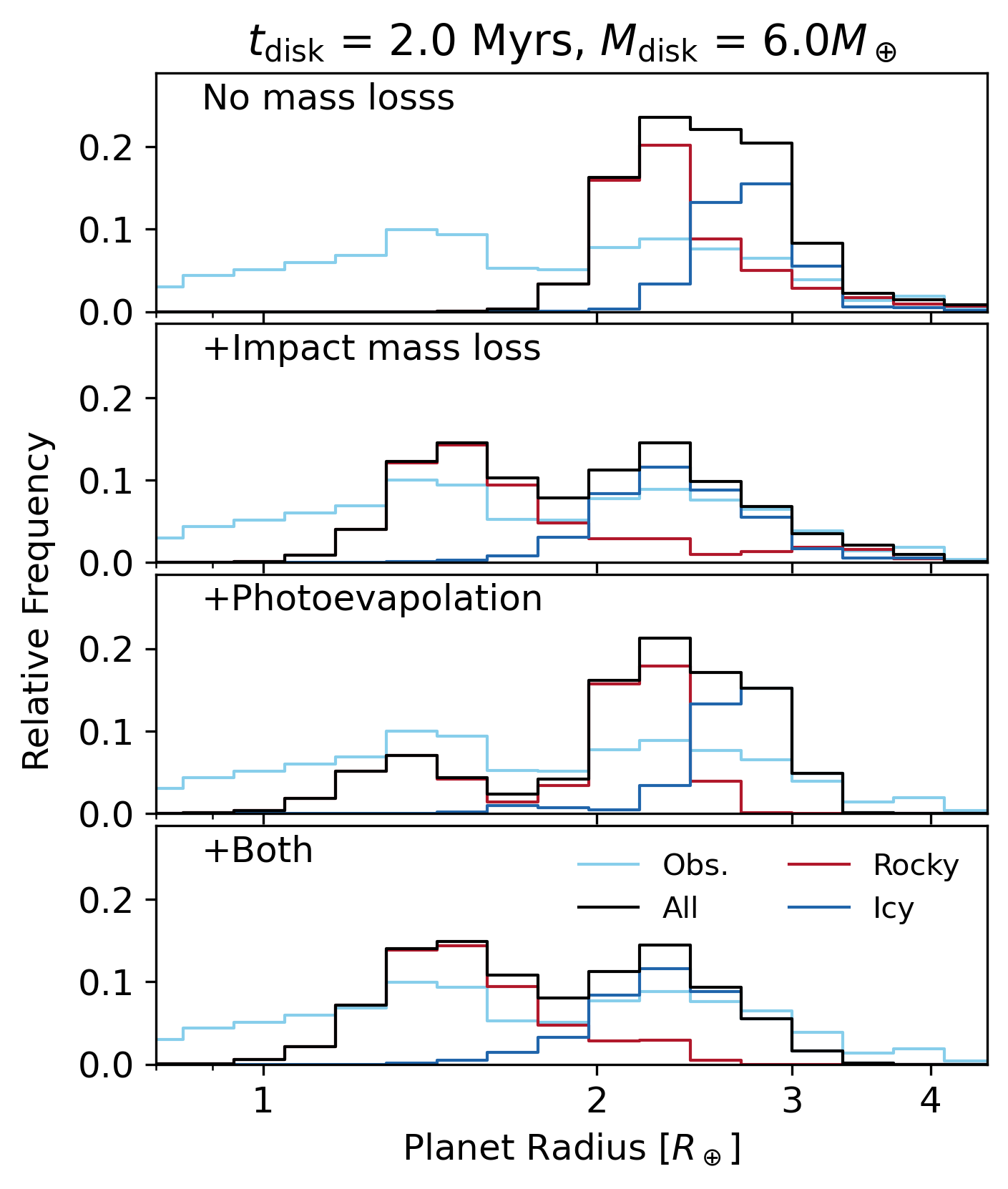}
    \caption{
    Radius distribution of planets obtained in {\it M6T3} ($M_\mathrm{disk}=6M_\oplus$, $t_\mathrm{disk}=3$ Myrs).
    We plot the cases without mass loss, with impact mass loss, with photoevaporation, and with both mass loss effects from top to bottom panels.
    }
    \label{fig: R_hist_MassLoss}
\end{minipage}
\begin{minipage}[b]{0.49\columnwidth}
    \centering
    \includegraphics[width=\linewidth]{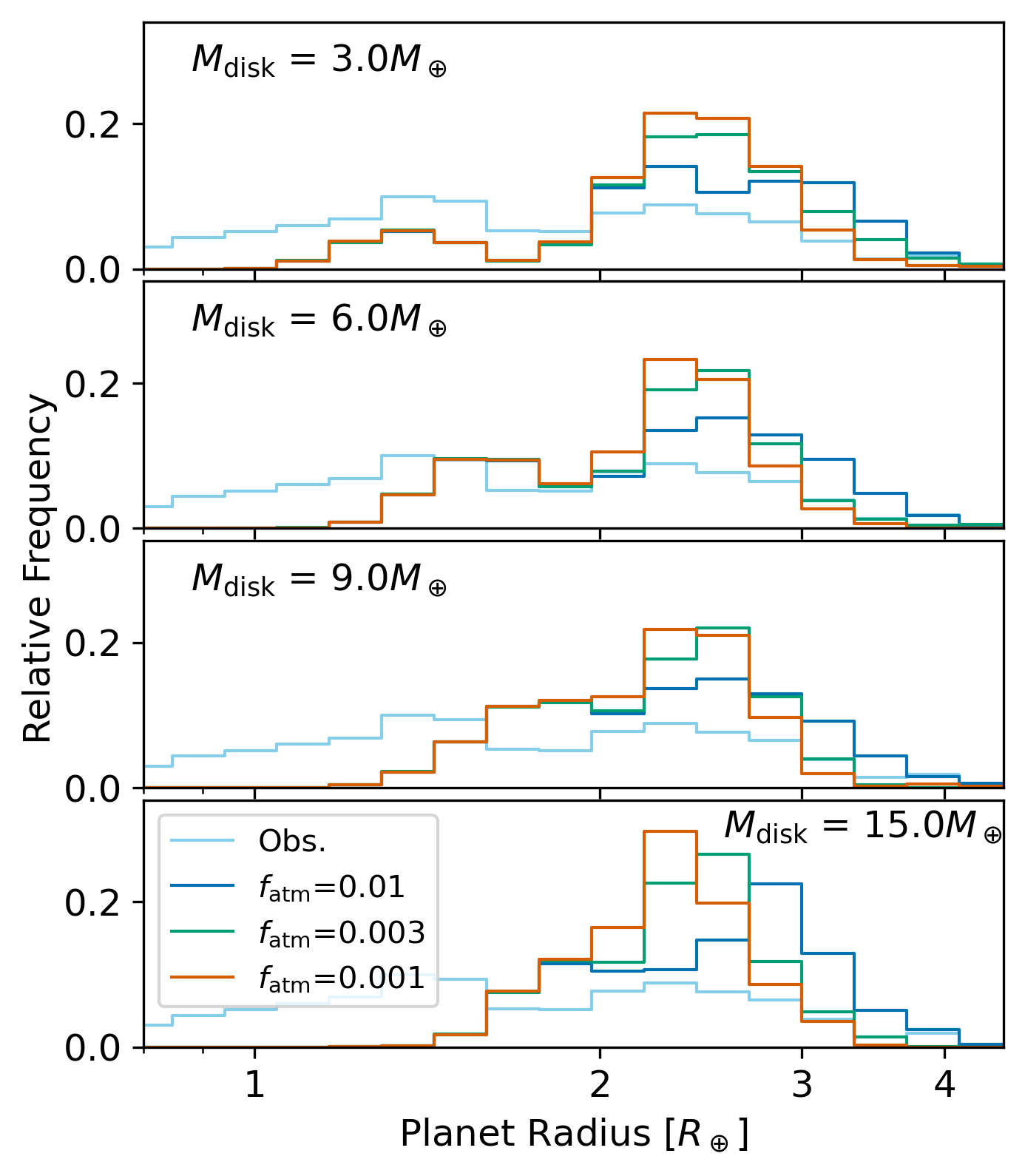}
    \caption{
    The histogram of the radius of planets with the different atmospheric mass fraction $f_\mathrm{atm}$.
    We show the simulations with $t_\mathrm{disk}=3$ Myrs.
    The planetesimal disk mass $M_\mathrm{disk}$ is labeled in each panel.
    The blue, green, and orange lines show $f_\mathrm{atm}=0.01$, 0.003, and 0.001, respectively.
    }
    \label{fig: Radius_Valley_3_fatm}
\end{minipage}
\end{figure}

We show the radius distribution of obtained planets in {\it M6T3} using different atmospheric mass loss models in Fig.~\ref{fig: R_hist_MassLoss}. Without any mass loss model, radius distribution shows the single peak around $2.4~R_\oplus$. The impact-induced mass loss and photoevapolation alter this radius distribution. The photoevaporation removes the atmosphere mainly from the rocky planets because rocky planets tend to be in inner orbit and receive stronger irradiation from the central star. On the other hand, impact-induced mass loss works on both rocky and icy planets in our model.

In this paper, we assume that the initial mass fraction of the atmosphere is $f_\mathrm{atm}=0.003$. We calculate the planetary radius using varying atmospheric mass fractions at the end of the gas disk phase and plot the radius distribution in Fig.~\ref{fig: Radius_Valley_3_fatm}. The figure indicates that the initial atmospheric mass fraction choice does not significantly affect the overall planet size distribution.

The above results indicate that the impact-induced mass loss plays an important role in our numerical model, independent of $f_\mathrm{atm}$. We assume that a giant impact with a mass ratio between the projectile and target $M_\mathrm{p}/M_\mathrm{t}>0.1$ completely removes the planetary atmosphere. This threshold value comes from the ratio of the heat capacities of the atmosphere and core obtained by \citet{Biersteker+2019}. Our model neglects the effects of the impact parameter and core composition, and it is unclear how the impact-induced mass loss changes with these parameters. Since impact-induced mass loss is the dominant mechanism for atmospheric mass loss, it should be investigated in future studies.

\section{Simulation of observational bias}\label{app: observational_bias}
A transit is only detectable if the planet’s orbital plane is sufficiently near to the line of sight between the observer and the star. We simulate transit observations of the planetary systems produced in our N-body simulations following  \citet{Izidoro+2017}. Each planetary system produced in our N-body simulations is observed with a given line of sight having a longitude $\theta$ and latitude $\phi$. 
We judge a planet transits the central star if
\begin{equation}
    \left| z_\mathrm{p}-z_\mathrm{v} \right|< b_\mathrm{crit} R_\mathrm{s}
\end{equation}
with
\begin{equation}
    z_\mathrm{p} = \frac{a \left(1-e^2\right)}{1+e\cos \left( \theta-\varpi \right)} \sin \left( \theta-\Omega \right) \sin I,
\end{equation}
\begin{equation}
    z_\mathrm{v} = \frac{a \left(1-e^2\right)}{1+e\cos \left( \theta-\varpi \right)} \sqrt{1- \sin^2 \left( \theta-\Omega \right) \sin^2 I} \tan \phi,
\end{equation}
where $b_\mathrm{crit}$ is the critical impact parameter set to 0.9, $R_\mathrm{s}$ is the radius of the central star set to the solar radius. $a$, $e$, $I$, $\Omega$, and $\varpi$ are semi-major axis, eccentricity, inclination, longitude of ascending node, and longitude of pericenter of the planet. We simulate the signal to noise ratio $\mathrm{SNR}$ following \citet{Weiss+2017} where
\begin{equation}
    \mathrm{SNR} = \frac{\left( R_\mathrm{p}/R_\mathrm{s} \right)^2 \sqrt{3.5 \mathrm{yr}/P}}{\mathrm{CDPP}_\mathrm{6hr} \sqrt{6 \mathrm{hr}/T_\mathrm{d}}}
\end{equation}
with
\begin{equation}
        T_\mathrm{d} = 13 \mathrm{hr} \left( \frac{P}{1\mathrm{yr}} \right)^{1/3} \sqrt{1-\left( \frac{\left| z_\mathrm{p}-z_\mathrm{v} \right|}{R_\mathrm{s}}\right)^2},
\end{equation}
where $\mathrm{CDPP}_\mathrm{6hr}$ is the 6-hour Combined Differential Photometric Precision. We randomly pick one of the $\mathrm{CDPP}_\mathrm{6hr}$ listed in Table 1 in \citet{Weiss+2017} for each transit. If $\mathrm{SNR}$ is larger than 10, we judge the transit is detected and store the orbital details of each detected planet. 
Repeating this manipulation for different $\theta$ (evenly spaced by 0.1 degrees from angles spanning from 30 to -30 degrees) and $\phi$ (evenly spaced by 1 degree and spanning from 0 to 360 degrees), we create a new set of observed planetary systems.  We use the output of our simulated observations to compare our results with observations.

\section{Example of Growth and Dynamical Evolution }\label{app: example_time_evolution}

\begin{figure*}[!ht]
    \centering
    \includegraphics[width=\linewidth]{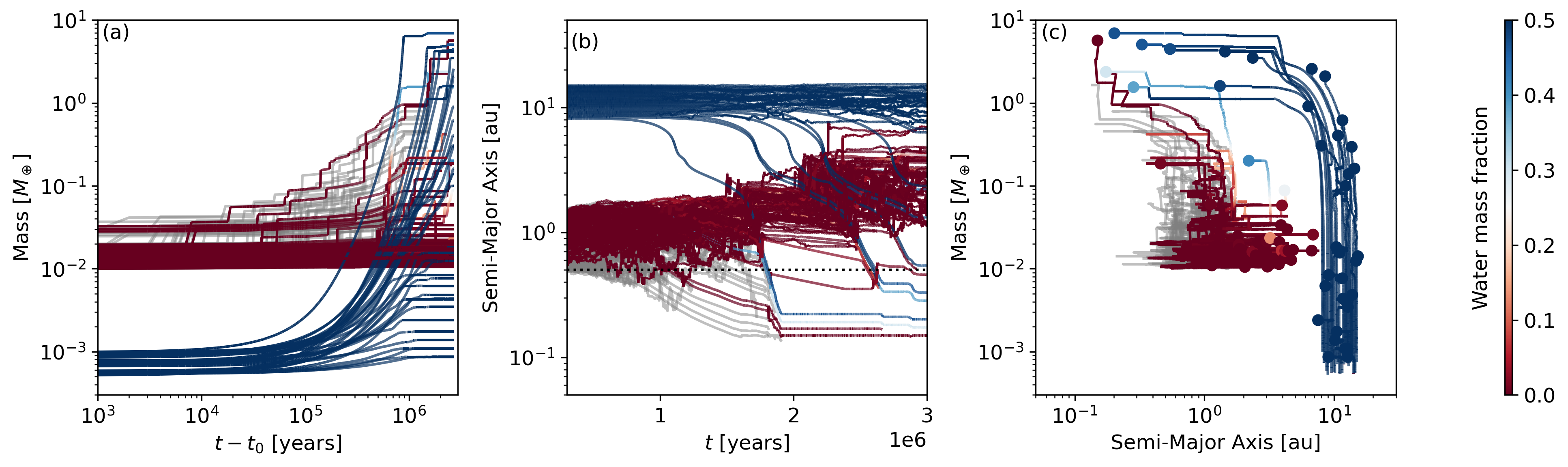}
    \caption{
    Representative example of the dynamical evolution of a planetary system during the gas disk phase.
    {\bf Left}: time evolution of planetary seeds' mass. Time is expressed relative to the start time of the simulation.
    Gray lines represent planetesimals accreted by other planetesimals by the end of the simulation. The gas disk lifetime is $t=2$ Myrs.
    {\bf Middle}: time evolution of planetesimals' semi-major axis.
    {\bf Right}: Growth tracks showing the evolution of our planetary objects by the end of the simulation ($t=50$ Myrs). In all panels, the color coding shows the water mass fraction of each planet. 
    }
    \label{fig: time_orbs_for_one_param}
\end{figure*}

In Figure~\ref{fig: time_orbs_for_one_param}, we show the time evolution of one planetary system produced in our {\it M6T2} simulations. We show the time evolution of mass and semi-major axis in the left and middle panels of Fig.~\ref{fig: time_orbs_for_one_param}. The color-coding corresponds to the water mass fraction of each planetesimal.
The right panel of Fig.~\ref{fig: time_orbs_for_one_param} shows the growth evolution tracks. This plot corresponds to the same simulation of Fig.~\ref{fig: Porb_Mp_for_one_param} in the main text.

\section{Kolmogorov-Smirnov test with our simulation data}\label{app: test_mixing}
\subsection{Kolmogorov–Smirnov statistics}

\begin{figure}
    \centering
    \includegraphics[width=0.5\linewidth]{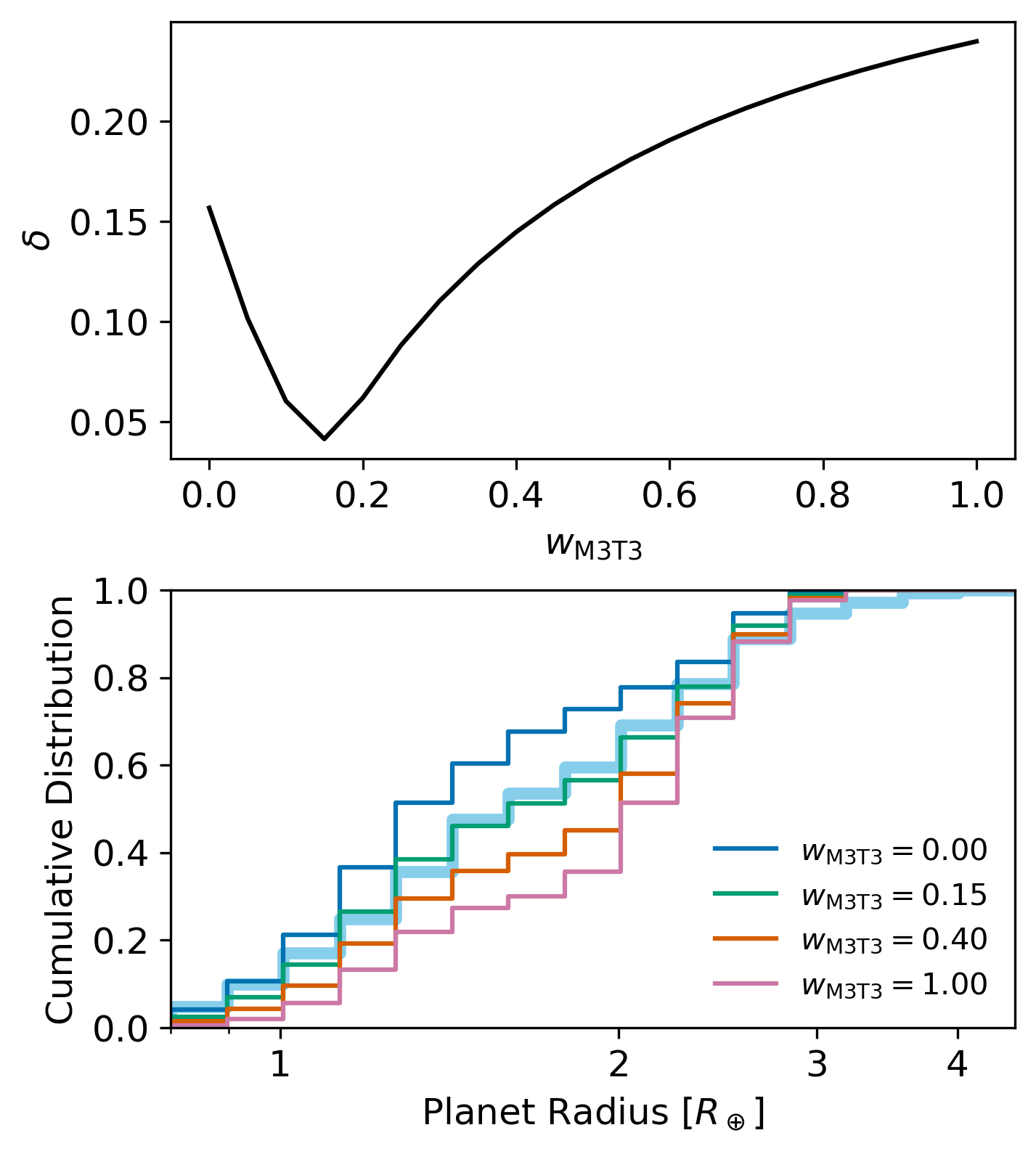}
    \caption{
    {\bf Upper panel}: Distance between the radius distribution of our mixed simulations and observations as a faction of the mixing ratio $w_\mathrm{M3T3}$. We mix {\it M3T2} and {\it M3T3} simulations.  {\bf Lower panel}: Corresponding radius distributions of mixed simulations with different mixing ratios $w_\mathrm{M3T3}$.
    }
    \label{fig:mix_test}
\end{figure}

Kolmogorov–Smirnov statistics $\delta$ is calculated by
\begin{equation}
    \delta = \max_i^n \left| F_\mathrm{sim}(i)-F_\mathrm{obs}(i) \right|,
\end{equation}
where $F_\mathrm{sim}$ and $F_\mathrm{obs}$ are normalized cumulative distributions of simulations and observations, respectively. $i$ is the index of each bin, and $n$ is the total number of bins. Figure~\ref{fig:mix_test} shows an example of how $\delta$ changes with the mixing ratio, where we mix the radius distribution obtained in the {\it M3T2} and {\it M3T3} simulations. We mix our simulated planetary systems considering relative mixing ratios $w_\mathrm{MxTx}$ ({\it MxTx} is the id of simulations) varying from 0 to 100\%, using steps of 5\%. For each possible combination of mixing ratios, we generate the corresponding $F_\mathrm{sim}$ and calculate $\delta$. The top panel shows $\delta$ for different mixing fractions $w_\mathrm{M3T3}$. The bottom panel shows the distributions for four mixing ratios $w_{M3T3}$. The minimum value of $\delta$ takes place when we mix the planetary systems from {\it M3T3} and {\it M3T2} considering $w_\mathrm{M3T3}=0.15$.

\subsection{The p-value calculation}
When we simulate transit observations of our systems from many lines of sight, we generate many observed systems. Some systems appear multiple times within this population of observed systems because each simulated planet system may be successfully observed from several lines of sight (see Appendix \ref{app: observational_bias} for details on our simulated observation method). Equivalently, our full set of final observed systems can be seen as a large dataset of ``observed values,'' where ``observed values'' refers to parameters such as planet radius, period ratio of adjacent planet-pairs, radius ratio of adjacent planet-pairs, and system multiplicity. We calculate the KS p-values for our datasets and real observations; however, we do not use the entire simulated dataset, as its sample size is artificially inflated from the original size. Instead, we compare the real observed population with a randomly selected sub-sample of our simulated observations~\citep{Izidoro+2017, Izidoro+2021a}.

We set the size of this sub-sample by first calculating the average number of observed planets and the average number of observed planet-pairs in the best-fit mixing solution. We found that 1.48 planets are observed on average in each of our simulated planetary systems. The average number of observed planet pairs in our simulations is about 0.48 per system because many systems produce single-planet systems. For each observed value, we define the effective sample size as $n_\mathrm{sim}=\Bar{N}_\mathrm{system} \times 50$, where $\Bar{N}_\mathrm{system}$ is the mean number of the observed value. 50 is the number of simulations that we have performed for each scenario. For the radius, $n_\mathrm{sim}$ becomes $74$. For the period ratio and radius ratio of adjacent planet pairs, $n_\mathrm{sim}=24$. For the multiplicity parameter, we use $n_\mathrm{sim} = 50$.

We next create a sub-sample population of size $n_\mathrm{sim}$ from the entire population. We randomly choose $n_\mathrm{sim}$ observed values, with replacement (meaning we can select the same data point multiple times). Our selected sub-sample population is weighted by $w_\mathrm{MxTx}$ as our best-fit mixing solution. 
We calculate the p-value of the KS test for the real observation and the obtained sub-sample. We repeat this re-sampling process 1,000 times and take the mean p-value as the p-value of the best-fit mixing solution.



\section{Individual distributions}\label{app: all_hist}

\begin{figure*}
    \centering
    \includegraphics[width=\linewidth]{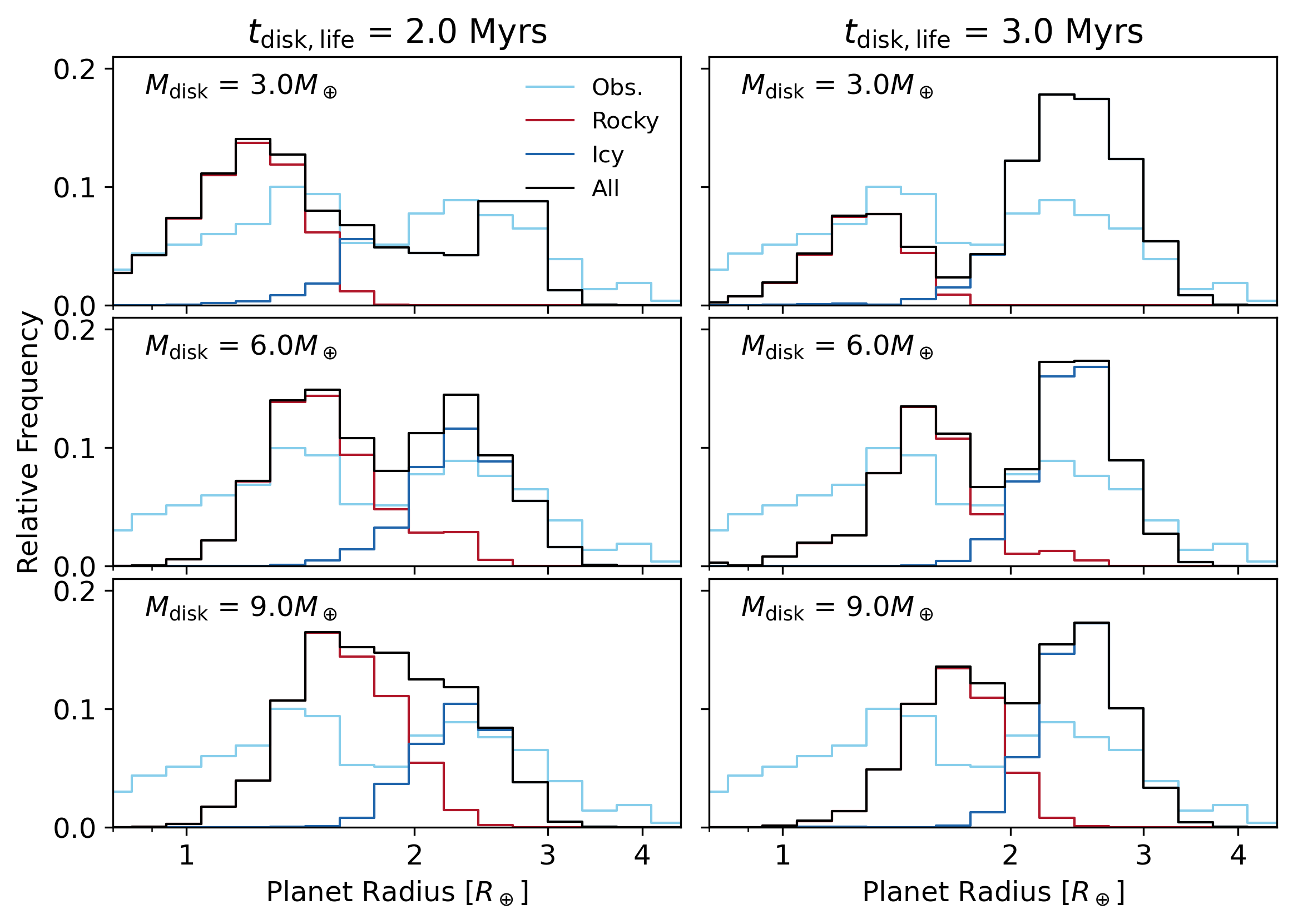}
    \caption{
    Radius distribution. The left and right columns show simulations with $t_\mathrm{disk}=2$ Myrs and 3 Myrs, respectively.
    The respective inner ring mass $M_\mathrm{disk}$ is shown in each panel.
    The black line shows the outcome of our simulations, including observational bias.
    The thin red and blue solid lines show the distributions of planets grouped into rocky and icy categories.
    The  blue solid line shows the CKS sample.
    }
    \label{fig: Radiut_Valley_all}
\end{figure*}

\begin{figure*}
    \centering
    \includegraphics[width=\linewidth]{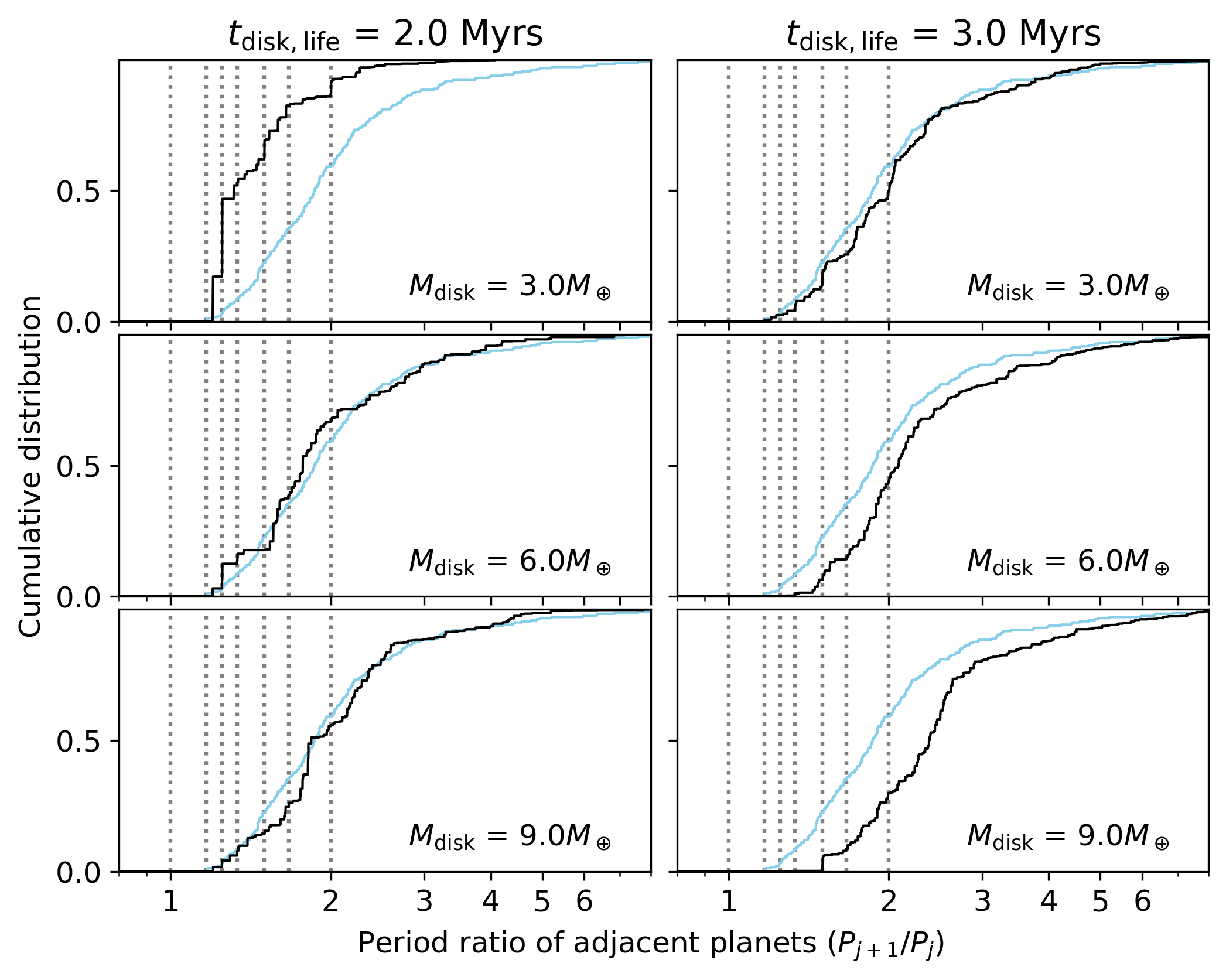}
    \caption{
    Period ratio of adjacent planets. The left and right columns show simulations with $t_\mathrm{disk}=2$ Myrs and 3 Myrs, respectively.
    The planetesimal disk mass $M_\mathrm{disk}$ is labeled in each panel.
    The black line shows the planets obtained in our formation model with the observational bias.
    The blue solid and gray dashed lines show the observation of CKS planets and all Kepler samples.
    }
    \label{fig: Period_Ratio}
\end{figure*}

\begin{figure*}
    \centering
    \includegraphics[width=\linewidth]{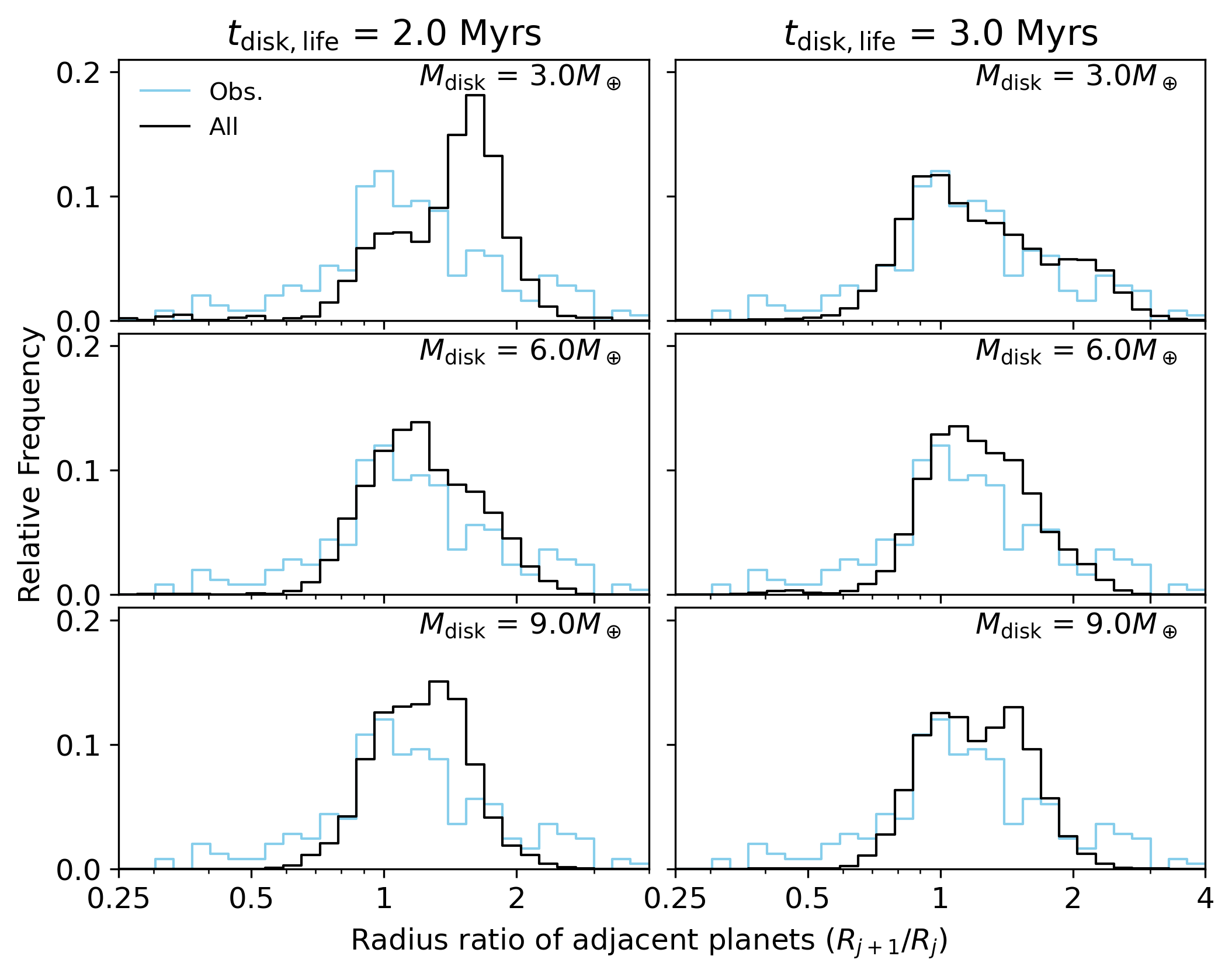}
    \caption{
    Radius ratio of adjacent planets.
    The respective inner ring mass $M_\mathrm{disk}$ is shown in each panel.
    The black line shows the outcome of our simulations, including observational bias.
    The thin red and blue solid lines show the distributions of planets grouped into rocky and icy categories.
    The  blue solid line shows the CKS sample.
    }
    \label{fig: Radius_Ratio}
\end{figure*}

\begin{figure*}
    \centering
    \includegraphics[width=\linewidth]{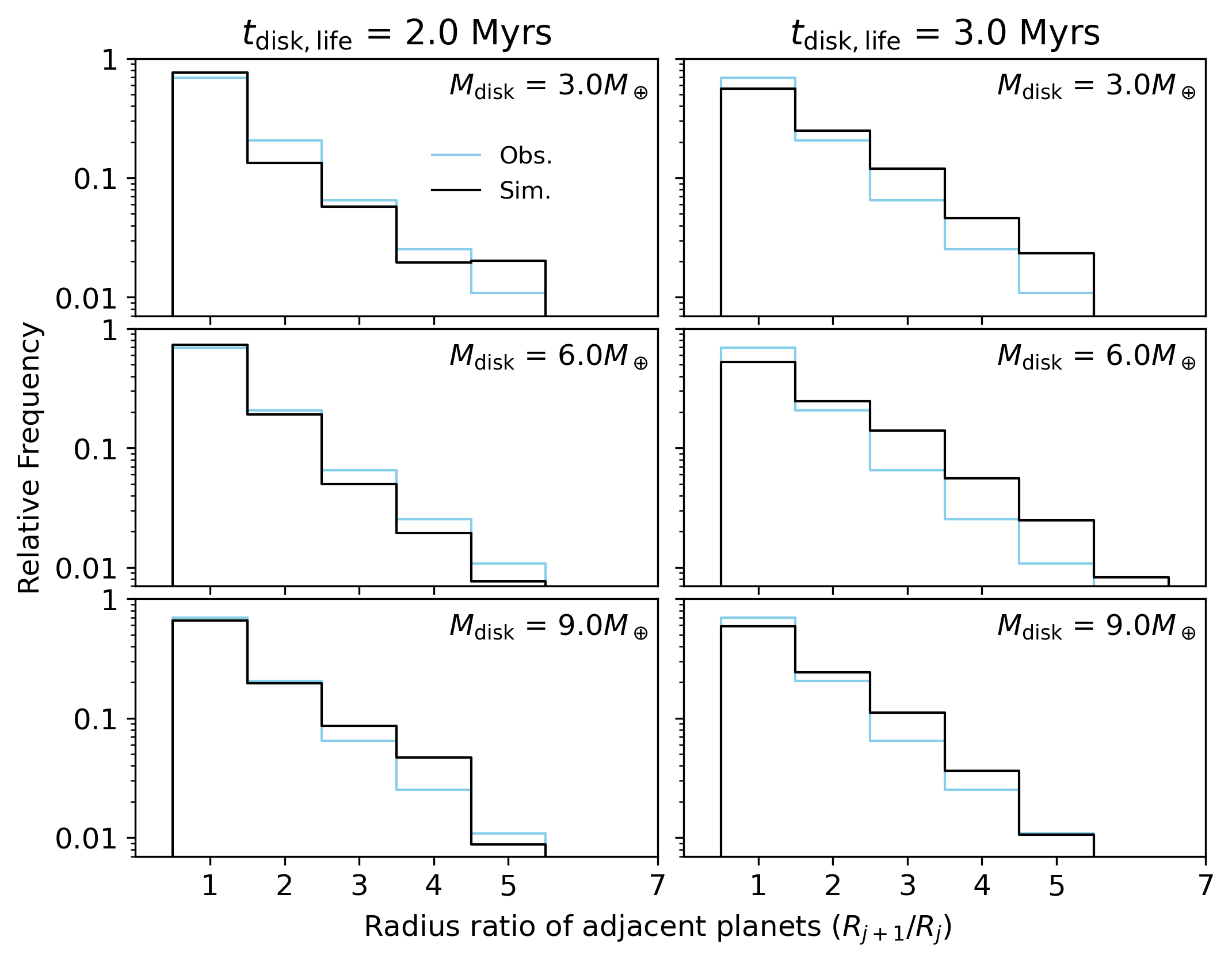}
    \caption{
     Observed number of planets.
    The respective inner ring mass $M_\mathrm{disk}$ is shown in each panel.
    The black line shows the outcome of our simulations, including observational bias.
    The thin red and blue solid lines show the distributions of planets grouped into rocky and icy categories.
    The blue solid line shows the CKS sample.
    }
    \label{fig: Multiplicity}
\end{figure*}

 In the main text we mix our different ring scenarios when comparing our results and observations. Here we separately present the radius distribution (Fig.~\ref{fig: Radiut_Valley_all}), period ratio distribution (Fig.~\ref{fig: Period_Ratio}), radius ratio distribution (Fig.~\ref{fig: Radius_Ratio}), and multiplicity of planetary systems (Fig.~\ref{fig: Multiplicity}) for each of our ring scenarios.  We do not show the results for $M_\mathrm{disk}=15 M_\oplus$ because we stopped these simulations at $t=20$ Myrs instead of 50 Myr.

\section{Final structure of planetary systems}

\begin{figure*}
    \centering
    \includegraphics[width=\linewidth]{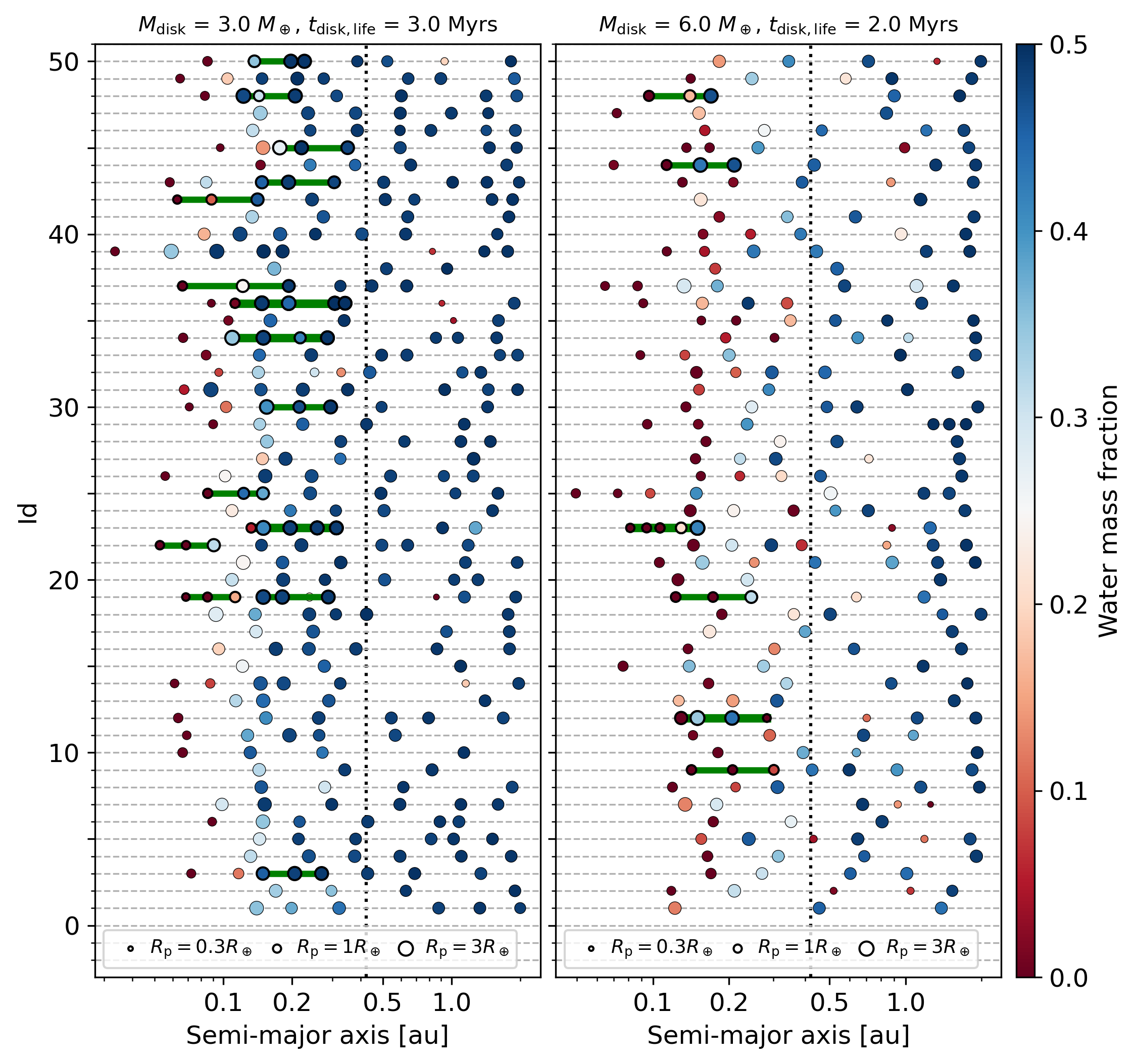}
    \caption{
    Final planetary systems produced in our  {\it M3T3} (left panel) and {\it M6T2} (right panel) simulations.
    The point's size scales with the planet's radius, and the color corresponds to its water mass fraction.
    The green line connects planets in mean motion resonances. The vertical dotted line corresponds to the location where the orbital period is $P=100\dy$.
    }
    \label{fig: Systems}
\end{figure*}

We show the final system architecture obtained in our numerical simulations of {\it M3T3} and {\it M6T2}. Figure \ref{fig: Systems} shows that the innermost planets tend to be rocky, and planets at larger orbital separations (r$>$0.5~au; P$>$100 days) tend to be water-rich. Out of our 100 simulations in {\it M3T3} and {\it M6T2}, five systems (id=19, 35 on the left panel and id=7, 23, 45 on the right panel of Figure \ref{fig: Systems}) formed a planet at around 1 au with water content lower than $5\%$ and mass larger than 0.1$M_{\oplus}$ (Mars-mass).  If we constrain our search to planets larger than 2 Mars' mass, the number of such systems decreases to two.



\bibliography{refs}{}
\bibliographystyle{aasjournal}



\end{document}